%% file: ms.tex
\pdfoutput=1
\documentclass[12pt,preprint]{aastex}
\usepackage{graphicx,apjfonts,emulateapj5,onecolfloat5,dblfloatfix}
\usepackage{amssymb,mathrsfs,color,gensymb}
\usepackage{longtable}
\usepackage{multicol}
\usepackage{blindtext}
\usepackage{ulem}
\usepackage{amsmath} 
\newfont{\rsfsten}{rsfs10 scaled 1200}
\newfont{\rsfsseven}{rsfs7 scaled 1200}
\newfont{\rsfsfive}{rsfs5 scaled 1200}

\slugcomment{Draft: \today, started March 12, 2020}

\shorttitle{}
\shortauthors{Zhao, Morris, Goss}
\begin{document}
\twocolumn[
\title{A population of compact radio variables and transients in the radio bright zone at the Galactic center observed
with the Jansky Very Large Array}  
\author{
Jun-Hui Zhao\altaffilmark{1}, 
Mark R. Morris\altaffilmark{2} \& 
W. M. Goss\altaffilmark{3}
} 
\affil{$^1$Center for Astrophysics | Harvard-Smithsonian, 60
Garden Street, Cambridge, MA 02138, USA; jzhao@cfa.harvard.edu}
\affil{$^2$Department of Physics and Astronomy, University of California Los Angeles, Los Angeles, CA 90095}
\affil{$^3$NRAO, P.O. Box O, Socorro, NM 87801, USA}
\begin{abstract}
Using JVLA data obtained from high-resolution observations at 5.5 GHz at multiple epochs in 2014 and 2019,
we have detected a population of radio variables and transients in the radio bright zone at the Galactic center.  
With observations covering a sky area of 180 arcmin$^2$ at an angular resolution of 0.4 arcsec, we report 
new detections of 110 Galactic center compact radio (GCCR) sources with a size of $<1$ arcsec. The 
flux densities of GCCRs exceed 70 $\mu$Jy, with at least 10$\sigma$ significance. Among these sources, 
82 are variable or transient and 28 are non-variable. About 10\% of them are expected to be extragalactic background sources.   
We discuss the possible astrophysical nature of the detected sources. 
As compared to the Galactic disk (GD) population of normal pulsars (NPs) and millisecond pulsars (MSPs), 
a majority (80\%) of the GCCRs appears to fall within the high flux-density tail of 
the pulsar distribution, as extrapolated from a sample of NPs in 
the Galactic disk. However, MSPs extrapolated from the GD population are too weak to have 
contributed significantly to the GCCR population that have been detected.
We also cross-correlated the GCCRs with X-ray sources in Chandra X-ray catalogs and 
found that 42 GCCRs have candidate X-ray counterparts.
Most of the GCCRs having X-ray counterparts  are likely to be 
associated with unresolved or slightly resolved radio jets launched from X-ray binaries with a compact 
object, either a black hole or a neutron star.
\vskip 5pt
\noindent 
{{\it Unified Astronomy Thesaurus concepts:} {\color{black}  
Center of the Milky Way; [Galactic center (565)]; Interstellar medium (847);
Radio continuum emission (1340); Black holes (162); Radio pulsars (1353); 
Millisecond pulsars (1062); Neutron stars (1108); White dwarf stars (1799); 
Discrete radio sources (389); Radio transient sources (1358); Radio interferometry (1346) 
}}
\end{abstract}
]

\section{Introduction}

The central parsecs of our Galaxy host a nuclear star cluster (NSC) with
a mass of 2-3 $\times10^7$ M$\odot$ \citep{sch2014,fel2014}. The mechanism of the
formation of the NSC is not clear among the two possible scenarios that
have been discussed:  in-situ formation \citep{mil2004,aha2015} versus the migration of stars
from a more distant region into the central parsec via the process of dynamical
friction \citep{tre1975,ant2015,arc2017}. There is evidence that a large fraction of the
cluster stars are as old ($\ge$ 10 Gyr) as those in the inner Galactic Bar/Bulge \citep{sch2019},
although the young massive stars at the center of the Galactic NSC  \citep{sch2003,ghe2005,gen2010,lu2013}
demonstrate the occurence of  ongoing star formation. The high stellar density
at the Galactic center (GC) and the high density of compact X-ray sources there \citep{zhu2018}
suggest that the central parsecs likely host a large population of stellar
binary systems belonging to low-mass X-ray binaries (LMXBs) \citep{mun2005,rem2006}. The compact
components in LMXBs are likely associated with either stellar black holes
(BH-LMXBs) or  neutron stars (NS-LMXBs) \citep{hai2018,zhu2018}. Some of the X-ray variables 
found in the GC appear to be associated with the activities occuring in
BH-LMXBs \citep{deg2015,deg2010,hai2018}. Some of the NS-LMXBs are thought to
host ordinary or normal  pulsars (NPs) and millisecond pulsars (MSPs) \citep{pfa2004,wha2012}. Recently,
an excess of X-ray source counts at the GC has been found to be comparable in magnitude
to the excess determined in globular clusters \citep{mun2005,hag2017},
where a large population of NPs and MSPs have been
found\footnote{http://www.naic.edu/~pfreire/GCpsr.html}$^,$\footnote{https://www.atnf.csiro.au/research/pulsar/psrcat/ 
(ATNF Pulsar Catalog)}. Consequently, one might expect the GC to host a large population of NPs and MSPs.
In fact, very few have been found, leading to the well-known "missing pulsar problem." 
{\color{black}\citep[{e.g., }][]{kra2000,joh2006,mac2010,bat2011,wha2012,eat2013b,dex2014,mac2015,eat2015,raj2017,bow2018}.
}

Based on a deep Chandra X-ray survey, \cite{mun2004} suggested that the vast majority of the Galactic center
X-ray sources are cataclysmic variables (CVs). CVs are low-mass close binary systems consisting of a white dwarf (WD) 
as a primary, accreting materials lost from a Roche-lobe filling, late type companion star. 
In the magnetic type (mCVs),  the  WD primaries harbor strong magnetic fields and  produce 
a stand-off shock above the WD surface \citep{aiz1973} while the accretion flow along the magnetic field  
reaches supersonic velocities. The post-shock region is hot (kT$\sim$10-50 KeV) and cools via thermal Bremsstrahlung 
radiation in hard X-rays. The  hard X-ray surveys of the INTEGRAL/IBIS-ISGRI and Swift/BAT surprisingly 
detected 1600 sources above 20 KeV \citep{bir2016,oh2018}. The follow-up deep X-ray observations with Chandra, 
XMM-Newton and NuSTAR revealed, indeed, that a large population of intermediate polars (IP), a type of mCVs,  
dominates the hard X-ray emssion in the central 10 parcsec \citep{mun2004,hea2013,per2015,hai2016,hon2016}.

Jet outflows often arise  from dynamic interactions within accretion disks associated with BHs  
\citep[{\rm e.g.,}][]{sha1973},  pulsars and MSPs in some NS systems \citep{eij2018}, and perhaps CVs \citep{cop2020,bar2017}, 
producing radio emission. Thus, high-resolution observations at radio wavelengths can provide substantial data for
diagnosis of the activities in the accretion process surrounding these compact objects.
 
However, only a few relatively bright radio sources \citep[e.g.,][]{zha1992,eat2013a} have so far been 
detected in the radio during their outbursts. Because of improvements of the VLA in both hardware and 
software for wideband operation, the enhanced JVLA sensitivity has allowed us to identify a population 
of compact radio sources embedded within the extended emission of the radio bright zone (RBZ) 
within the Galaxy's central 15$'$, or 35 pc.

In addition, low-frequency emission from compact sources  is subject to scatter-broadening. However, 
the discovery of the magnetar, SGR J1745-29,  {\color{black} or PSR J1745-2900 hereafter}, 
located just 3"  from the bright compact radio source associated with the central black hole, 
Sgr A*, indicates that the effect of the scattering screen could be up to three orders of 
magnitude smaller than expected \citep{spi2014,bow2014}. {\color{black}While the temporal 
scatter-broadening of PSR J1745-2900 is less than expected by orders of magnitude,
the angular broadening is consistent with that of Sgr A*, suggesting they both
lie behind the same strong (angular) scattering screen. Therefore, lines of sight 
to Sgr A* are still strongly scattered in the image domain.}
Also, the radio counterpart of the X-ray Cannonball \citep{par2005}, a possible runaway neutron star 
from the Sgr A East supernova remnant (SNR), shows a peak intensity of 0.5 mJy beam$^{-1}$ at 5.5 GHz 
with a resolution of 1" while the surrounding pulsar wind nebula (PWN) becomes slightly resolved at 
a resolution of 0.5" with A-array data \citep{zmg2013,zmg2020}. The radio emission from MSPs is typically 
much weaker. Observations show that  MSPs are indeed weaker than  NPs, with mean values of logarithmic luminosity 
(Log S$_{\nu}$D$^2$ [mJy~kpc$^{-2}$]) of  0.5$\pm$02   for a sample of 31 MSPs located in the Galactic disk (GD)
as compared to 1.50$\pm$0.04 for 369 normal pulsars (NPs), where S$_{\nu}$ is the observed flux density at 1.4 GHz
and D is the  distance in kpc \citep{kra1998,lor1995,tay1993}. The spectra of MSPs are 
steep but comparable to those of NPs, with spectral index of $\alpha\sim-1.7$, where $S_\nu\propto\nu^\alpha$. 
Scaling the GD samples to the GC distance of 8 kpc, we expect mean values
of flux density at 5.5 GHz to be 5 and 50 $\mu$Jy at 5.5 GHz for the populations of  
MSPs and NPs, respectively. Such low flux-density values estimated for
both NPs and MSPs partially explain the difficulty in the detection of pulsars, especially of MSPs, 
at the GC. Given that the mean  flux density for MSPs at the GC distance is close to the VLA sensitivity limit, we only 
expect detection of a few candidates for bright MSPs with the present capability of the VLA. Most  
pulsars are polarized, with linear polarization of a few percent to 100\% at 1.4 GHz \citep{joh2018}. 
On the other hand, the radio emission from MSPs is expected to be steady and highly polarized at lower
frequencies, while weakly polarized or not polarized at $\nu>$ 3 GHz \citep{kra1999}.
The non-variable MSP emission gives us a handle for finding the faint emission from MSPs
in a high-dynamic-range image produced by combining VLA data observed at multiple epochs.

Based on our recent 5.5-GHz VLA observation in the A-array on 2019-9-8, along with two previous
observations on 2014-5-26 and 2014-5-17, we focus on searching for the relatively bright
compact radio sources outside both the HII complexes of Sgr A West and Sgr A East.
The HII gas in Sgr A West is associated with the circumnuclear disk, and the nearby complex 
HII regions in Sgr A East, denoted as A, B, C and D in \cite{goss85} or G-0.02-0.07 in \cite{mill11}, 
are associated with ongoing formation of high mass stars. 

The rest of paper is organized as follows:
Section 2 describes the observations, data reduction and imaging used for searching for compact radio sources.
Section 3, along with Appendix A, presents a catalog of 
the Galactic center compact radio (GCCR)
sources found in the RBZ from this search.
Section 4 presents identifications of X-ray counterparts.
Section 5 discusses the astrophysical implications of 
the GCCR sources and constraints regarding their nature,
and section 6 summarizes our conclusions.

\begin{table*}[]
\tablenum{1}
\setlength{\tabcolsep}{1.7mm}
\caption{Log of datasets and images}
\begin{tabular}{lccccclccc}
\hline\hline \\
&\multicolumn{5}{c}{\underline{~~~~~~~~~~~~~~~~~~~~~~~~~~~~~~~~~ UV data ~~~~~~~~~~~~~~~~~~~~~~~~~}}& 
&\multicolumn{3}{c}{\underline{~~~~~~~~~~~~~~~~~~~~~~~~~~~~ Images  ~~~~~~~~~~~~~~~~~~~~~~}} \\
{Project ID}&
{Array}&
{Band}&
{$\nu$}&{$\Delta\nu$~~~} &
{HA range} &Epoch & Weight & {\color{black}($\theta_{\rm maj}^{\rm FWHM}\times\theta_{\rm min}^{\rm FWHM}, 
{\rm PA}$)}& RMS\\
{} &
{} &
{} &
{(GHz)} &
{(GHz)} &  &
{(day)} &
{(R)}   &
{(arcsec$\times$arcsec, deg)}&
{($\mu$Jy beam$^{-1}$)}
\\
{(1)}&{(2)}&{(3)}&{(4)}&{(5)}&{(6)}&{(7)}&(8)&(9)&(10)\\
\hline \\
\vspace{5pt}
19A-289 &A &C$^\ddagger$& 5.5&2&$+0^h.3$ --- $+2^h.3$&2019-9-8&$-$0.25&0.62$\times$0.23, 14&7.5\\
\vspace{5pt} 
14A-346 &A &C$^\ddagger$& 5.5&2&$-0^h.5$ --- $+3^h.5$&2014-5-26&0&0.58$\times$0.26, 20&5.3\\
\vspace{5pt} 
14A-346 &A &C$^\ddagger$& 5.5&2&$-3^h.2$ --- $+0^h.7$&2014-5-17&0&0.59$\times$0.23, $-$15&5.8\\
\hline
\end{tabular}\\
\begin{tabular}{p{0.90\textwidth}}
{\footnotesize
(1) The JVLA program code of PI: Mark Morris.
(2) The Array configurations.
(3) The JVLA band code; "C" stands for the 5-GHz band.
(4) The observing frequencies at the observing band center.
(5) The bandwidth.
(6) The  hour angle (HA) range for the data.
(7) The date corresponding to the image epoch.
(8) The robustness weight parameter.
(9) The FWHM of the synthesized beam.
(10) The rms noise of the image.} 
{\footnotesize
$^\ddagger$Correlator setup: 64 channels in each of 16 subbands with channel width of 2 MHz. 
 }\\
\end{tabular}
\end{table*}

\section{Observations and data reduction}

Deep observations achieving a sensitivity  of  a few $\mu$Jy beam$^{-1}$ are enabled at
the JVLA by improvement in both hardware and software for  wideband capability. 
Radio detections now become possible of some stellar sources at the GC such as 
X-ray binaries and bright pulsars. Therefore, we can constrain their natures using X-ray, infrared and
follow-up radio observations. 


\subsection{Data sets \& calibrations}

New JVLA observations in the A configuration were carried out on 2019-9-8 at 5.5 GHz.
Along with two previous A-array observations at epochs 2014-5-26 and 2014-5-17, we have
a total of three A-array data sets at 5.5 GHz. These observations were all carried out
with an identical VLA standard correlator setup for wideband continuum covering 2 GHz bandwidth, with a
single field  pointing at a position near the geometrical center of the Sgr A East radio 
shell\footnote{RA(J2000)=17:45:42.718, Dec(J2000)=$-$29:00:17.97}.
Table 1 summarizes the three sets of uv data  (columns 1 - 7). 

The data reduction was carried out using the CASA\footnote{http://casa.nrao.edu} software package of the NRAO. 
The standard calibration procedure for JVLA continuum data was applied. J1733-1304 (NRAO 530) 
was used for complex gain calibrations. The flux-density scale was calibrated using 
standard calibrators, either 3C 286 (J1331+3030) and/or 3C 48 (J0137+3309). Corrections 
for the bandpass shape of each baseband and the delay across the 2 GHz bandwidth were determined 
based on the data from flux-density calibrators. The accuracy of the flux-density scale at 
the JVLA is 3\%$-$5\%, limited by the uncertainty of the flux density of the primary calibrator, Cygnus A \citep{per17}.

\begin{figure*}[!ht]
\includegraphics[angle=-90,width=230mm]{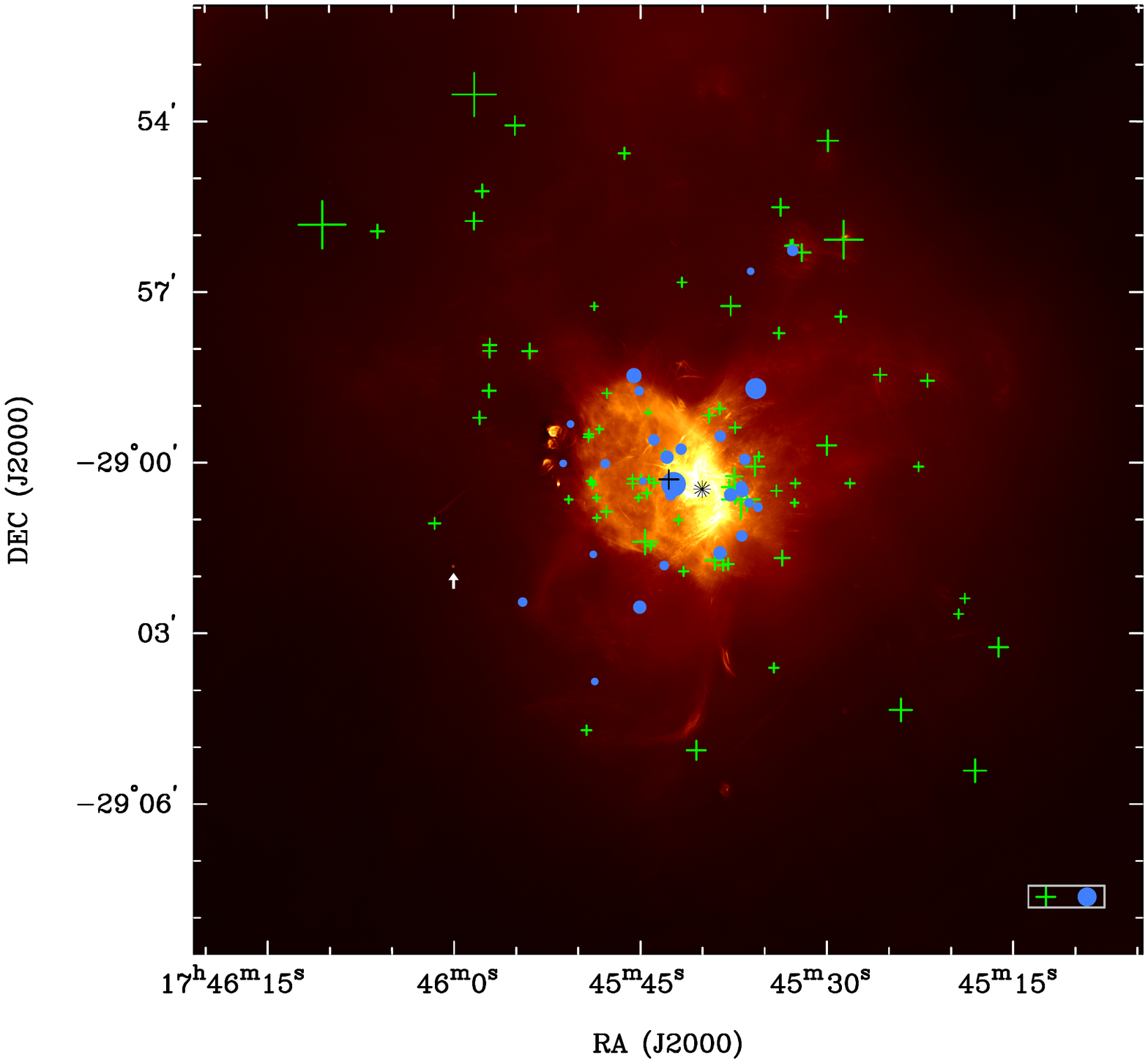}
\caption{The 5-GHz image of the radio bright zone (RBZ),
constructed with hybrid data obtained from a combination of observations
with the JVLA in the A, B and C arrays, the old VLA in D array, and the GBT
in single-dish mode, giving good uv coverage between 0 and 800 k$\lambda$.
The RMS noise is $\sim$2 $\mu$Jy beam$^{-1}$. The synthesized FWHM beam is 0.68"x0.47"
(8.4$^{\rm o}$). The green cross symbols, "+", mark the positions of 
the radio variables and transients (N=82) and the blue dots indicate the locations of  
non-variables (N=28). These GCCR sources are newly identified from the JVLA
high-resolution images observed at 5.5 GHz in A-array during 2014
May and 2019 September. A total of 118 compact sources are located outside
Sgr A West and the HII regions A, B, C, D in \cite{goss85} or G-0.02-0.07 in \cite{mill11}.
The size of the symbols is scaled as $\sim10" {\left[ S/1~{\rm mJy}\right] }^{1/3}$,
where $S$ is the source flux density. The size of the symbols in the box at 
bottom-right corresponds to a 1 mJy source for both variables ("+") and non-variables
(dot). The Galactic plane is oriented in this figure at a position angle of $\sim30^{\rm o}$.
A white arrow marks the location of the nebular source G-0.04-0.12 (see Figure 2 for details).
The black "+" marks the phase center of the data or the pointing center of the observations; 
the coordinates of the phase center
are given in the footnote$^3$ in the text. The black 16-pointed star marks the position of Sgr A*.}
\label{fig1}
\end{figure*}

\begin{figure*}[!h]
\centering
\includegraphics[angle=0,width=60mm]{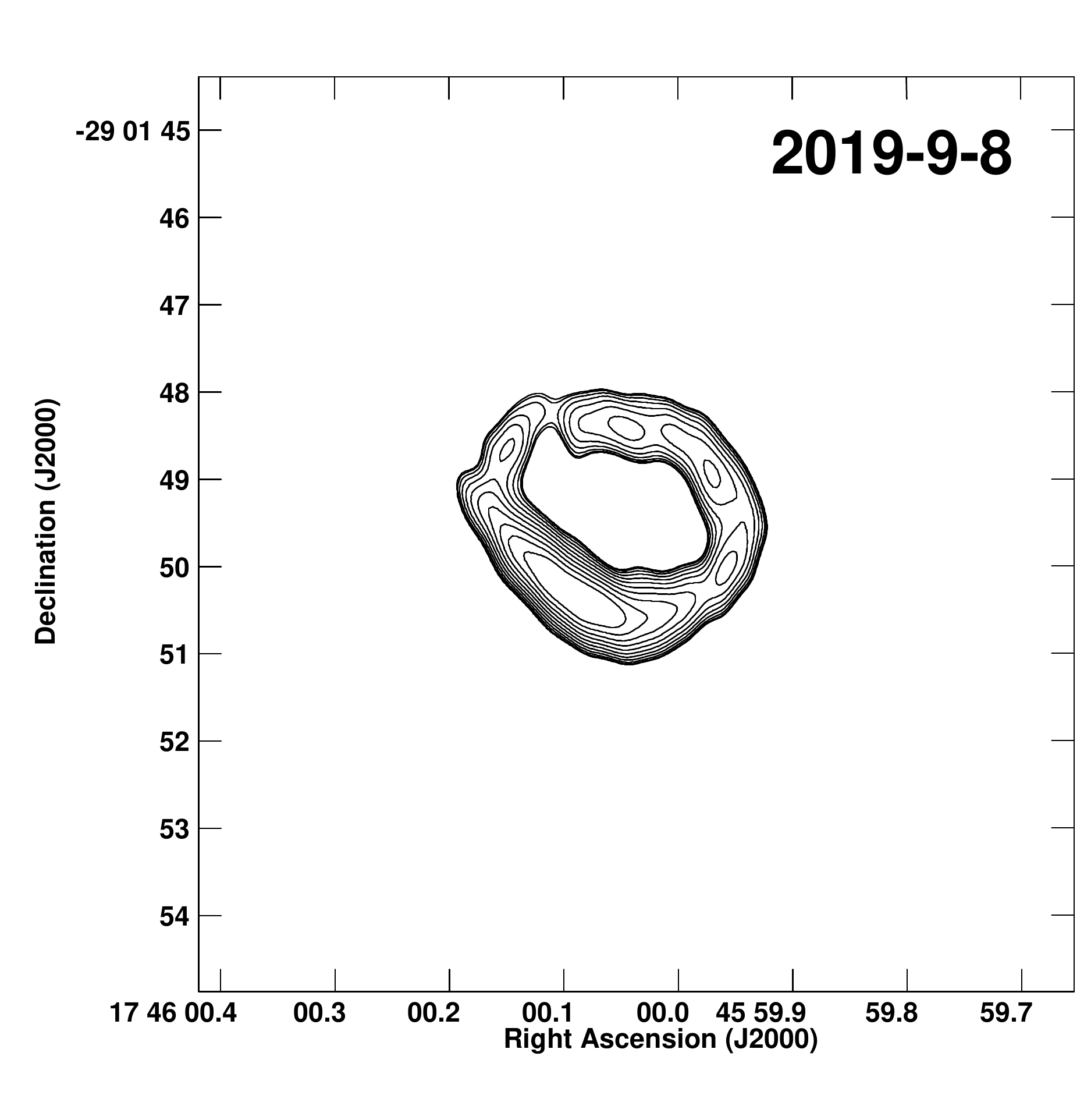}
\includegraphics[angle=0,width=60mm]{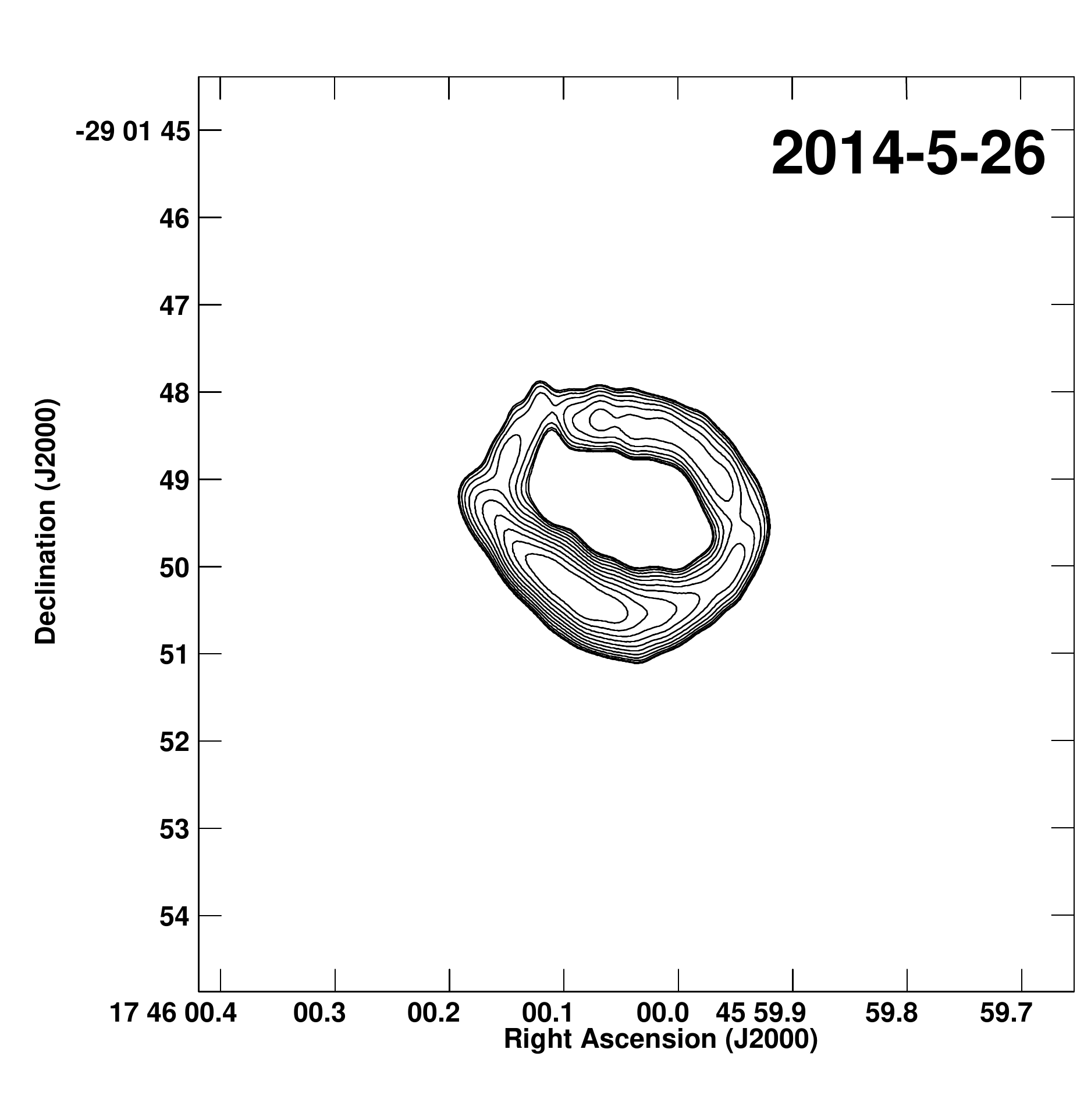}
\includegraphics[angle=0,width=60mm]{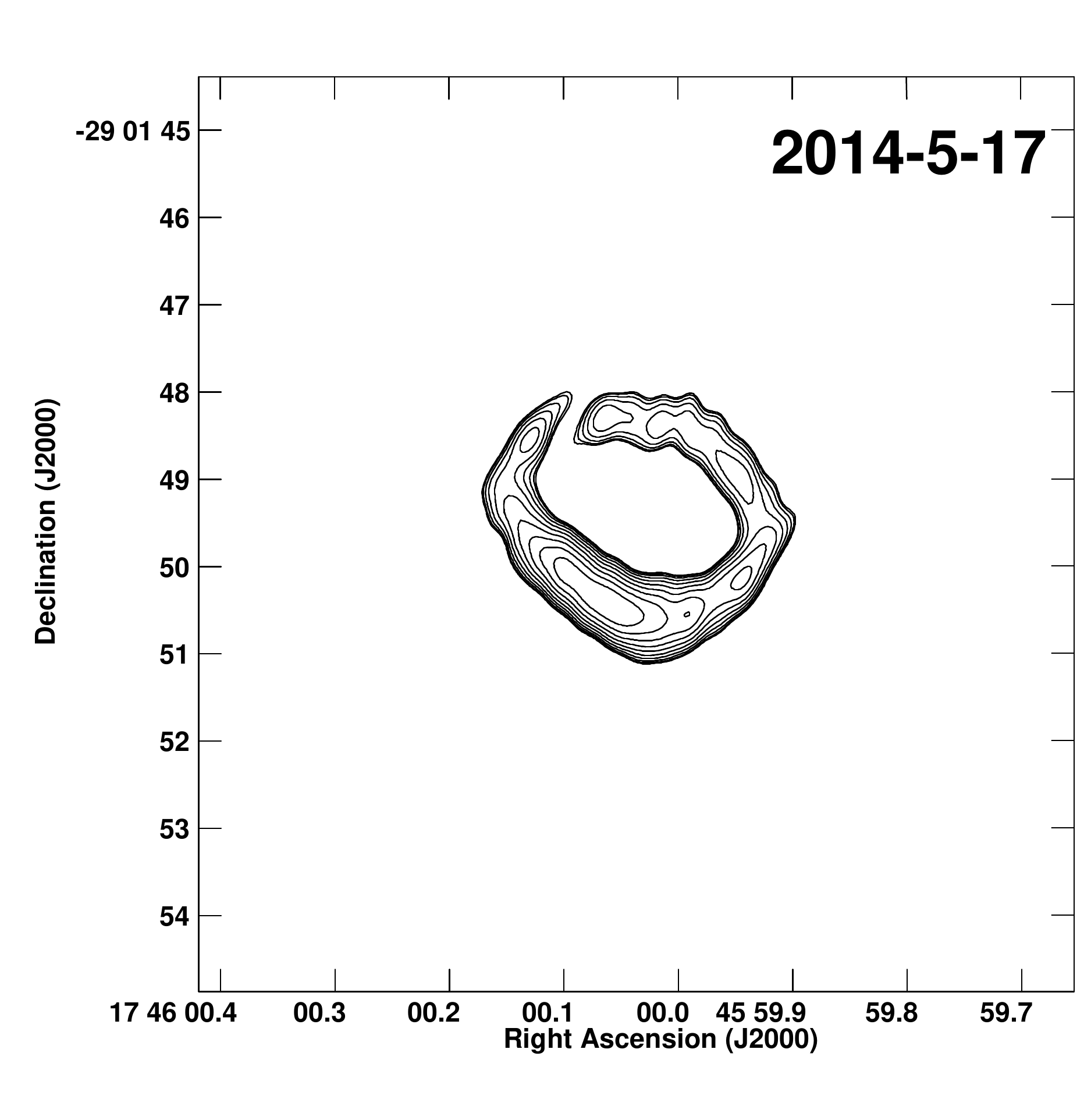}
\caption{The 5.5-GHz images of the  nebular source G-0.04-0.12  
\citep{mill11} produced by filtering out the short-spacing uv data at the three epoch's 
observations at 2019-9-8 (left), 2014-5-26 (middle) and 2014-5-17 (right)
and applying the correction for PB attenuation. The dirty images were cleaned with 
the MS-MSF algorithm \citep{rau11} and the images with cleaned components were finally  
convolved with a common beam of FWHM 0.50"$\times$0.24" ($-0.05^{\rm o}$) instead of 
their synthesized beams. The intensities of the source observed at the three epochs  
can be compared without bias. Then, the difference between 
the levels of background emission caused by the different HA coverages
of the data are correctable in the measurements of flux density (see Section 2.2.3).
The total flux density of an extended source can be determined reliably in all epochs.
The contours are 1$\sigma\times$(10, 11, 13, 16, 20, 25, 31, 38, 46, 55, 65), where
the local RMS noise  $\sigma=$10$\mu$Jy beam$^{-1}$. The integrated flux densities of
14.8$\pm$0.3, 14.6$\pm$0.3, and 14.7$\pm$0.3 mJy are determined for the epochs 2019-9-8, 2014-5-26
and 2014-5-17, respectively. The source G-0.04-0.12 is marked with a white arrow in the wide
field image (Figure  1),  located at a distance of 3.5 arcmin from the field center, close to the 
contour at the half power beam width (HPBW).
}
\label{fig2}
\end{figure*}

\subsection{Imaging}
Following the procedure for high dynamic range (DR) imaging that we developed recently
\citep{zmg2019} and applying it to the Sgr A data with CASA, we have produced a 
deep image of the GC RBZ at 5.5 GHz with  hybrid data obtained from a combination of 
observations with the JVLA in the A, B and C-arrays, the old VLA in D-array and the GBT in 
single-dish mode, providing  good uv coverage between 0 and 800 k$\lambda$ (see the background image
in Figure 1). The RMS noise in a region far  from  the bright emission region 
Sgr A West is 2 $\mu$Jy beam$^{-1}$. The ratio of the peak intensity, 0.8 Jy beam$^{-1}$, 
to the value of the RMS noise implies a DR of 400,000:1. Indeed, the rms noise is similar to 
the mean 5.5 GHz flux density of MSPs at the Galactic center, as extrapolated from the 1.5 GHz 
value assuming a frequency dependence of $\sim\nu^{-1.7}$. 

However, the Sgr A West region that hosts the nuclear star cluster emits a diffuse continuum  
with a total flux density of  $\sim$ 15 Jy at 5 GHz, distributed in the circumnuclear disk (CND), 
in addition to the prominent mini-spiral feature \citep{ekers83}. The confidence level for 
detections of weak compact sources near a strong radio complex may be compromised because of 
various issues in sampling and imaging radio interferometer array data. In a study of 
compact sources lying within a large field covered by a single primary beam (PB), both 
PB attenuation and smearing effects due to both bandwidth- and time-averaging can produce 
a loss in intensity of a compact source. Corrections must be applied for the errors caused 
by these effects. 

\subsubsection{Contamination from short-spacing power}
High-amplitude short-spacing visibilities produce  confusion due to the corresponding extended emission.
In particular, an extended structure sampled by short baselines in high resolution imaging 
can emerge from the analysis of several  small clumps that potentially lead to confusion in 
the identifications of weak compact sources. In addition, the relatively weak emission from 
compact radio sources is easily hidden in a bright extended emission complex, such as the 
Sgr A complex (see Figure 1). The high radio power at the Galactic center may explain
why only a few bright compact sources in the RBZ have so far been reported. Furthermore, owing to limitations of
the available deconvolution algorithms, false compact sources may be produced by residual 
sidelobes of a dirty beam near a strong extended emission region. For example, the FT of 
a uniform disk is a 2D Airy function. A dirty image of a VLA sampled disk source is difficult 
to clean because of strong sidelobes and residual phase errors \citep[{\rm e.g.,}][]{led92}. 
To quantitatively evaluate the contamination from  
residual sidelobes of a disk source, we carried out simulations with CASA processing of 
visibility models using the same procedure  as utilized for the real data. Three visibility 
data sets for a model  of the diffuse disk of Sgr A West (75"$\times$40", PA=0$^{\rm o}$) were made,
corresponding to the uv coverages sampled in each of the three epochs' observations (Table 1). 
We cleaned the sidelobes of the disk model with the CASA task 
{\sf TCLEAN}, and noticed that compact clumps present outside the disk do mimic compact radio 
sources up to an intensity of 0.2 mJy beam$^{-1}$. These compact clumps are the sidelobes of the 
discrete sampling function simulated for the disk model, but appear as discrete radio sources 
due to the limitation in the clean process for a disk of emission. The limitation of 
handling the sidelobes from a complex emission source can therefore produce false compact sources. 

One way to resolve this issue is to process the imaging  with a cut-off of the short-baseline 
data that corresponds to extended emission. With the VLA A-array data at 5.5 GHz, we find that 
using only the longer baseline ($>100~k\lambda$) data, corresponding to sampling the small scale 
($<2$ arcsec) emission, work well for diminishing the level of the residual sidelobes. Following 
the same procedure described above, we cleaned the disk model with lower baseline cutoff of 100~k$\lambda$. 
The RMS outside the disk in the cleaned image is improved by a factor of 15 as compared to that with all 
the A-array data; the maximum of the surrounding clumps drops by a factor of 100, and 
the RMS is reduced to a level of 1 $\mu$Jy beam$^{-1}$.

This algorithm has been applied to the real data. With the three A-array data sets, 
we constructed images having 20k$\times$20k pixels covering the 15'$\times$15' of the RBZ region
using only  the longer baseline uv data ($>100~k\lambda$). The properties of the high-resolution 
images at the three epochs are summarized in  columns 8-10 of Table 1.
A nebular source G-0.04-0.12 {\color{black}(Figure 2)}  
with a size of 3"$\times$4",  presumably with a constant flux density, is
located southeast of Sgr A East \citep{mill11}. After filtering out the short baseline data, the resultant image 
is used to verify the consistency of the flux-density scale using our method. The images 
made from  the longer baseline data ($>100~k\lambda$) at the three epochs show a nearly 
identical ring of the nebula, demonstrating  consistent images obtained  with the algorithm
discussed here. We find no suspected arifacts surrounding the nebular ring in the cleaned images 
down to a level of 10 $\sigma$.

\subsubsection{PB-corrections and uncertainty}

With the sensitivity of the JVLA, we are able to detect a compact source at a large radial distance from 
the telescope pointing center. In a region far from the telescope pointing center, the uncertainty in
the correction for attenuation becomes large. We carried out PB corrections with AIPS task 
{\sf PBCOR} using a polynomial model updated by \cite{per16}:

\begin{equation}
\centering
A(x)=\sum_{i=0}^{i=3} A_{2i}x^{2i}
\end{equation}
with a variable $x=\nu r$, where $\nu$ is the observing frequency 
and $r$ is the angular distance from the PB center,
and $A_{2i}$ is the polynomial coefficients used in fitting the VLA PB.
We corrected the image to the 2\% level of the PB.  At a large distance 
from the PB center, the corrections are subject to an increased uncertainty. 
The uncertainty $\sigma_A$ of $A(x)$ can be assessed with the formula:

\begin{equation}
\centering
\sigma_A =\sqrt{\sum_{i=0}^{i=3} \Big(\frac{\partial A(x)}{\partial A_{2i}} \sigma_{A_{2i}}\Big)^2}, 
\end{equation}
where $\sigma_{A_{2i}}$ are the uncertainties in the polynomial coefficients
$A_0$, $A_2$, $A_4$ and $A_6$ given in \cite{per16}.

\renewcommand{\thefigure}{3}
\begin{figure*}[!h]
\label{fig:3}
\centering
\includegraphics[angle=0,width=185mm]{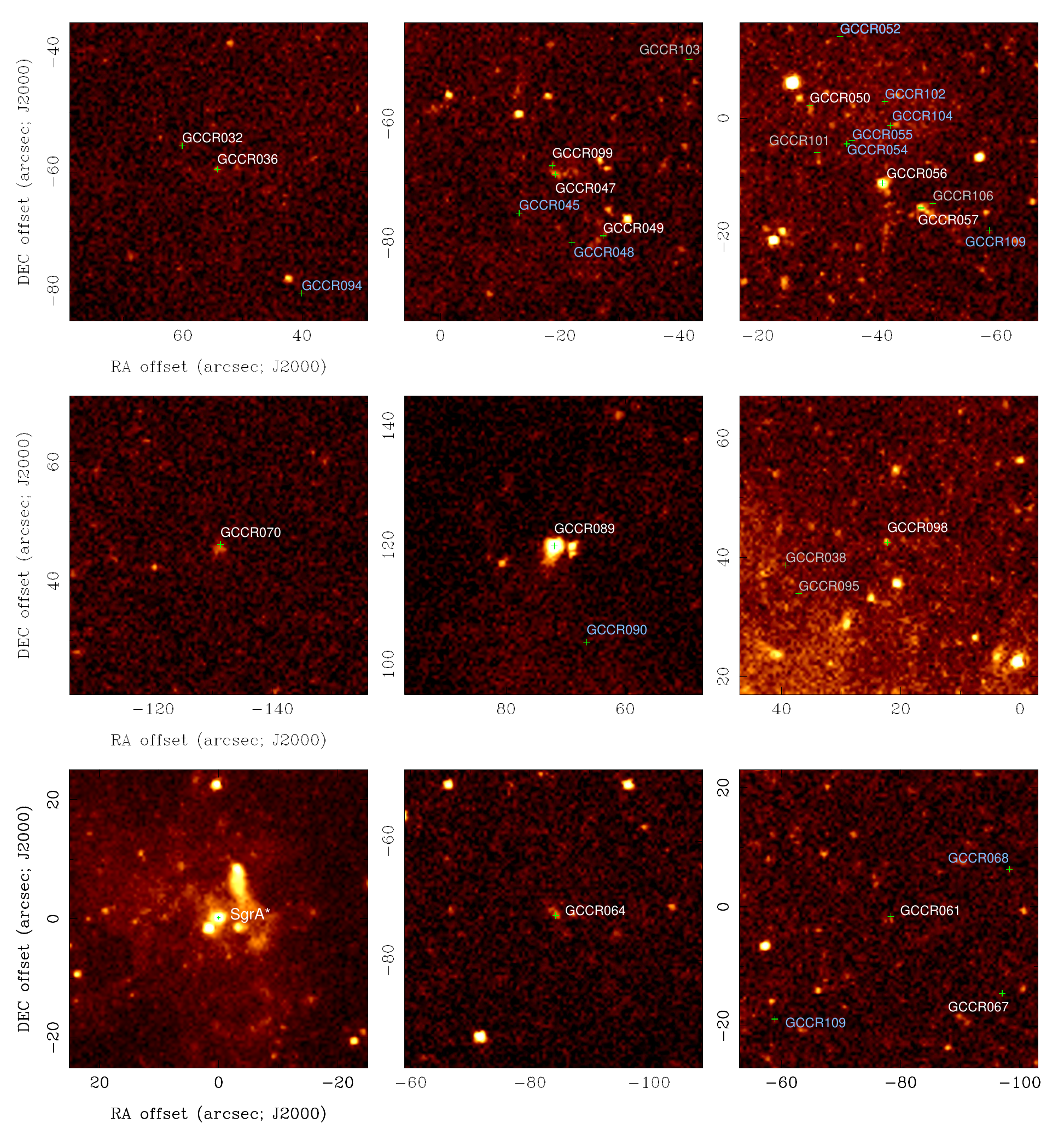}
\caption{\scriptsize{X-ray images showing the X-ray counterparts to the cores
of the selected GCCR soures.
The background X-ray image,  from \cite{zhu2018}, has a spatial resolution of 0.5".
The "+" symbols mark the positions of the GCCR sources. The bottom-left panel is the field including Sgr A* 
and the central panel is the field including the X-ray cannonball (GCCR089).
Sgr A* and the cannnonball are both compact and bright in X-rays and radio,
and these two compact sources were used to align the coordinate frames between
the X-ray and radio images. The coordinates on all nine fields are  
the angular offsets in RA and Dec from Sgr A*. {\color{black}The colors represent the angular offsets between 
a GCCR and its possible X-ray counterpart in the ranges of $<1"$ (white), $1"-2"$ (grey), and
$>2"$ (light blue), corresponding to identification code "y", "y?" and "n" marked in column 13 of Table 2.} 
}\label{fig:3}}
\end{figure*}

\subsubsection{Hour-angle vs. variation of flux density}

The variability in flux density is one of the properties that facilitates differentiating between
various types of compact radio sources \citep[{e.g.,}][]{kra2006,bro2018,cor2011}. Often, 
a compact radio source is associated  with an extended emission feature surrounding an 
unresolved core.  In such cases, the combination of intrinsic structure of a source and 
HA-range in uv sampling may produce a false variability. To access the uncertainty introduced 
by such an effect, we simulated a linear source described by a 2D Gaussian function 
(0.8"$\times$0.1", 0$^{\rm o}$ or 90$^{\rm o}$ ) of 0.5 mJy by adding an unresolved core, 
or a point source, of 0.1 mJy at the center of the linear source. With two intrinsic PA values 
of 0$^{\rm o}$ and 90$^{\rm o}$ for the linear components, ten models of simulated linear+core 
source were distributed at ten positions at radial distances of 
up to 2 arcmin from the phase center to simulate three data sets. The simulated uv data sets 
were made by sampling the source models in the A-array configuration with HA-coverage 
identical to the real data, producing 128 channels covering a 2 GHz bandwidth at 5.5 GHz.
Then, following the same setup and procedure as used to process the real data, the simulated 
data sets were Fourier-transformed by averaging every 2 channels, with the lower uv cutoff of $100$ k$\lambda$; 
and the dirty images were 
cleaned with CASA programs. With the AIPS task {\sf JMFIT}, we made a Gaussian fit to 
the model sources found in the cleaned images, and find that the loss in flux density 
is in the range between 2\%-10\% of the input values for the extended linear feature.
The loss in peak intensity is larger for the point source, or the core, falling in the range between
30\% to 65\% of the input values due to the bandwidth smearing (BWS) effect\footnote{This effect 
is proportional to $\frac{\delta \nu}{\nu_0}$, the ratio of channel width to the central 
observing frequency, and to $r_\theta$, the angular distance of a source to the phase center
of the interferometer array \citep{tms2017}.}. The loss in peak intensity of the cores due to
the BWS effect is correctable with {\sf JMFIT}. For example, the correction factor $\sqrt{1+\beta^2}$  
can be computed, where $\beta$ is provided in Eq(A4) of  Appendix A. 

The images of the nebula G-0.04-0.12 at three epochs made with three A-array data sets  
(Figure 2) were used to examine the issues of  flux-density variation caused by changes 
of HA coverage. The difference in HA coverage between the three epochs' observations does 
cause a minor difference in the level of a shallow negative hole underlying and surrounding 
an extended emission feature although the same uv cutoff ($>100$ k$\lambda$) was consistently 
applied. The apparent flux densities from the positive HA images (2019-9-8 and 2014-5-26 of 
Figure 2) agree well with each other while the apparent flux density derived from the 
2014-5-17 image corresponding to the data taken with a negative HA-coverage decreases 
significantly due to a relatively deeper shallow negative area surrounding the source. 
The apparent flux densities integrated over the source are 14.6$\pm$0.2 mJy, 14.5$\pm$0.2 
mJy and 11.3$\pm$0.20 mJy determined from the images of 2019-9-8, 2014-5-26, and 2014-5-17, 
respectively. The corresponding values of the flux density contributed from the shallow negative 
hole underlying the source are $-0.2\pm$0.2 mJy, $-0.1\pm$0.2 mJy, and $-3.4\pm0.2$ mJy. 
The zero or background level biased by the HA coverage in the flux-density measurements can be 
corrected by simply subtracting the negative flux density from the apparent source flux density.  
After corrections for the local negative level, the variation in the final reported flux 
densities of 14.8$\pm$0.3 mJy, 14.6$\pm$0.3 mJy and 14.7$\pm$0.3 mJy at the three epochs 
for the nebula are consistent with the RMS fluctuations at a level of less than 2\%, similar 
to the uncertainties propagated from the flux-density calibrations. In summary, the analysis of 
the nebular data verifies that a signficant difference in the zero level surrounding a 
source is potentially present due to differences in HA coverage for the uv data,
but the bias in the determination of source flux density with Gaussian fitting is 
correctable with subtraction of a fitted background level using the AIPS task {\sf JMFIT} automatically.  
Our examinations of G-0.04-0.12 images provide the procedure used for reliable measurements 
of the compact sources that are discussed in the rest of the paper.   

Finally, we assessed a possible loss in  source intensity caused by time-average smearing (TAS), 
using a model of circular uv-coverage with Gaussian tapering \citep{bri99}. We find that 
the fractional losses due to TAS for the sources listed in Table 2 are less than 2\% in general.
For the sources located within the HPBW of the PB, the loss is less than 0.5\%. Therefore, 
no corrections for the effect of TAS have been  applied.

\section{Catalog of compact radio sources}

A population of compact radio sources within the RBZ -- at a level down to tens of $\mu$Jy --
has been revealed with our 5.5 GHz VLA observations \citep{zmg2020}. The sub-mJy compact 
radio sources are thought to consist of a mixture of thermal sources associated with 
compact/ultra-compact HII regions and non-thermal synchrotron sources that are related 
to the particle acceleration occuring in the accretion process associated with closely 
interacting binary stars or perhaps with isolated pulsars and PWNs. In this paper, we 
primarily searched for the GCCRs outside of  the known  HII regions. From the VLA A-array 
images observed in the three epochs, we have identified 110 compact sources located outside 
Sgr A West and the Sgr A East HII regions, G-0.02-0.07, but within a radius of $7.5'$ from 
the pointing center of the observations. The search criteria for the GCCRs are based on their
compactness (a size of $\theta_{\rm maj}<1$") and  significance ($S/\sigma>10$).

Appendix A discusses the GCCR catalog (Table 2) in detail, along with the presentation of 
high-resolution  images of every GCCR source at 5.5 GHz (Figure A1).

\section{X-ray counterparts}

\subsection{Catalogs of X-ray sources and Chandra images}
Catalogs of X-ray sources in the region surrounding the Galactic center have been produced 
using data from the Chandra X-Ray Observatory by combining observations taken on many 
different occasions \citep{mun2003,mun2006,mun2009,zhu2018}. The central Chandra pointing with the ACIS-I 
detector covers an area of 17'$\times$17', which is comparable to the  
primary beam of the  JVLA at 5.5 GHz. The Chandra field overlaps strongly with our JVLA field of 
view\footnote{The Zhu et al. (2018) field center is at RA(J2000) = 17:45:40.044, 
Dec(J2000) = $-$29:00:28.04, which is  displaced by 36.40"  from the JVLA pointing center. }.
We therefore cross-correlated our GCCR catalog with the ultra-deep point-source X-ray catalog of 
\cite{zhu2018}, which incorporates Chandra observations between 1999 and 2013, and 
reports 3619 sources in the 2-8 keV band within 500" of Sgr A*.   For regions outside  
the \cite{zhu2018} catalog area, we used  earlier catalogs covering a greater area 
\citep{mun2008,mun2009} for the cross-correlation analysis.

For candidate X-ray counterparts to GCCR sources, we also carried out a careful 
examination of the X-ray image used by \cite{zhu2018} to construct their catalog. 
The two previously known compact X-ray and radio sources --- Sgr A* and the Cannonball 
--- were used to align the coordinate frames of the X-ray and radio images. The reference 
centers of both images were shifted to the position of Sgr A*: RA(J2000) = 17:45:40.0409, 
Dec(J2000)= -29:00:28.118. The precision  in the positional alignment between the Chandra 
X-ray and JVLA images is $\sim0.25(S/\sigma)^{-1}$ arcsec ($< 0.1$ arcsec), where 
${S}/{\sigma}$ is the ratio of  signal to noise for the reference sources used in 
the alignment. We found candidates using the catalog cross-correlation, and then used 
the images to verify the coincidence and to look for possible structure in the X-ray 
morphology that might be helpful in assessing the correspondence. Figure 3 plots 
examples of those possible X-ray counterparts in 50"$\times$50" sub-frames used in the 
identification process. We examined as well the three Chandra images in the 2-3.3, 3.3-4.7  
and 4.7-8 keV bands for the central 900"\footnote{https://chandra.harvard.edu/photo/2010/sgra/  
where the FITS images of these bands were obtained.} that fully cover the RBZ observed at 5.5 GHz 
for the distribution of the GCCRs. Thus, the cross-correlation analysis between X-ray and 
5.5 GHz radio is spatially complete. Identifications of X-ray counterparts to individual GCCRs 
are tabulated in Table 3. Column 1 is the GCCR-ID. Column 2 lists  the name of an X-ray source 
in the Chandra X-ray Observatory  catalog,  CXOGC\#,  where \# stands for truncated J2000 
coordinates of the source JHHMMSS.S-DDMMSS \citep{mun2009}. In the diffuse X-ray source
catalog of \cite{mun2008},  the name of an X-ray source is denoted as G\# where \# stands 
for DDD.DDD$\pm$D.DDD, the Galactic coordinates in degrees. Column 3 gives the source sequential 
numbers (SS\#) in the deep X-ray catalog of \cite{zhu2018}. Column 4 gives the angular offsets 
between the GCCRs and their X-ray counterparts. Column 5 lists the 2-8 keV photon flux ($S_{\rm 2-8~keV}$)
reported in the catalogs, or the range of reported fluxes (upper --- lower values) 
\citep{mun2008,mun2009,zhu2018}. Column 6 gives the references from which the
X-ray data are used in the identifications.

\begin{table}[!h]
\footnotesize
\centering
\tablenum{3}
\setlength{\tabcolsep}{1.mm}
\caption{Identifications of candidate X-ray counterparts of GCCRs}
\begin{tabular} {ccrcccr}
\hline\hline \\
{GCCR-ID}&
{CXOGC\# or G\#}&
{SS\#}&
{$\Delta\theta_{\rm X-R}$}&
{$S_{\rm 2-8~keV}$}&
{Ref.$^\ddagger$} \\
&&&
(arcsec)&
(10$^{-7}$ ph cm$^{-2}$ s$^{-1}$)&
 \\
(1)&(2)&(3)&(4)&(5)&(6)\\
\hline\\
GCCR001 &J174610.5-285550& \dots &1.5&60 & a \\
GCCR004 &J174558.4-285546& \dots &1.2&3.6 & a \\
GCCR014 &J174549.3-290442& \dots &0.6&5.8 & a \\
GCCR015 &J174549.3-290442& \dots &0.5&5.8 & a \\
GCCR031 &J174544.9-290017& \dots &0.8&5.5 & a \\
GCCR032 &\dots          & 2625  &0.2&2.8 & b\\
GCCR035 &J174544.2-290018& \dots &0.7&7.5 & a \\
GCCR036 &J174544.1-290128& 2594  &0.5&4.5& a,b \\
GCCR037 &J174543.9-290020& \dots &1.5&15& a \\
GCCR038 &\dots          & 2502  &2 &2.1&  b\\
GCCR041 &J174541.5-285651& \dots &1.8&1.7& a \\
GCCR047 &G359.925-0.051 & \dots &\dots      &18.1& c \\
GCCR049 &\dots          & 1898  &1         &8.12 --- 7.50& b \\
GCCR050 &\dots          & 1875  &1         &7.4 --- 5.6& b \\
GCCR056 &J174536.9-290039& 1767  &0.2       &43.0 --- 42.2& a,b \\
GCCR057 &G359.933-0.037  & \dots &\dots      &22.8& c \\
GCCR058 &G359.941-0.029  & \dots &\dots      &18.7& c \\
GCCR059 &J174535.6-285953 & 1617  &1.6&6.44 --- 3.60 & a,b \\
GCCR061 &J174534.0-290030 & 1429  &0.8       &4.96 --- 4.80 & a,b \\
GCCR064 &J174533.5-290140 & 1375  &0.4       &9.10 --- 6.17 & a,b \\
GCCR067 &J174532.6-290043 & 1250  &1         &1.28 --- 1.20 & a,b \\
GCCR070 &J174530.0-285942 & 956   &0.7&17.5 --- 17.0 & a,b \\
GCCR072 &J174528.8-285726 & 852   &0.4       &4.1 --- 3.5 & a,b \\
GCCR073 &J174528.6-285605 & 819   &0.8       &9.6 --- 8.7 & a,b \\
GCCR074 &J174528.1-290021 & 774   &1  &3.34 --- 1.50 & a,b \\
GCCR077 &\dots           & 389   &1  &1.12& b \\
GCCR082 &J174516.1-290315 & 185   &0.6&40.3 --- 30.0   &a,b \\
GCCR085 &J174550.6-285919 & 3042  &0.3       &3.2 --- 1.7  &a,b \\
GCCR087 &J174548.7-290350 & 2926  &0.4 &3.2 --- 2.4  &a,b \\
GCCR089 &J174545.5-285828 &\dots  &0.3       &170&d,a \\
GCCR092 &J174544.6-290020 &\dots  &2  &2.7&a\\
GCCR095 &\dots           &2477   &1.9&4.23&b \\
GCCR096 &J174542.5-290033 &\dots  &1.2&1.9&a \\
GCCR097 &J174542.2-290024 &\dots  &1.9&3.1&a \\
GCCR098 &J174541.7-285945 & 2369  &0.2       &6.80 --- 6.63   &a \\
GCCR099 &G359.925-0.051  &\dots &\dots       &18.1&a \\
GCCR100 &J174538.6-285933 &\dots &2   &1.2&a \\
GCCR101 &J174537.6-290035 &1857  &1.6 &10 --- 6.0 &a,b \\
GCCR103 &J174536.8-290117 &\dots &0.4        &1.5&a \\
GCCR106 &G359.933-0.037  &\dots &\dots        &22.8&c\\
GCCR107 &J174536.1-285638 & 1671 &0.5        &190 --- 186 &a,b \\
GCCR110 &J174532.7-285617 & 1263 &0.7        &9.4 --- 6.9 &a,b \\
\hline
\end{tabular}
{\footnotesize
$^\ddagger$References:  a. \cite{mun2009}; b. \cite{zhu2018},
c. \cite{mun2008}; and d. \cite{par2005}.
 }\\
\end{table}

In addition, notes for those GGCRs involving extended X-ray emission sources such as halos and elongated
nebulae, or possessing possible IR identifications, are given in Section 4.2.

In short, a total of 42 GCCRs have candidate X-ray counterparts; most of them (27) have a positional 
offset between X-ray and radio, $\Delta\theta_{\rm X-R}$, less than 1" or less than twice the Chandra 
resolution; the rest of them (15) have $\Delta\theta_{\rm X-R}=$1" to 2". 
The probability that a GCCR source has an accidental coincidence within 1" or 2",
given the number of $\sim$3900 reported X-ray sources \citep{mun2009,zhu2018} lying within the area covered by our radio survey, is
1.9\% or 7.7\%, respectively. 
Thus, most of the 42 GCCRs with candidate X-ray counterparts are likely related to the X-ray sources. 
The majority of the GCCRs (68) do not have X-ray counterparts within 2 arcsec. The presence or absence of 
a candidate X-ray identification for the indvidual GCCRs is also indicated in Table 2 in the Appendix.
\subsection{Notes to the X-ray counterparts}
{
\parindent 0pt \hangindent 0pt \vskip 5pt
${\bf GCCR001}~-$ CXOGC J174610.5-285550 \citep{mun2009}, is offset by 1.5" 
from the radio source. An extended X-ray halo of  size 15" surrounds
the compact X-ray source in the 2.0-3.3 keV and 3.3-4.7 keV bands, 
but no significant X-ray emission is present in the 4.7-8 keV band.

\parindent 0pt \hangindent 0pt \vskip 5pt
${\bf GCCR047}~-$ This radio source is associated with a bright spot in an X-ray complex,
see the top-middle panel in Figure 3. The X-ray source is listed in \cite{mun2008} as G359.925-0.051 with
a power-law spectrum of $\Gamma=1.77$ and X-ray luminosity of 3$\times10^{32}$ erg s$^{-1}$; it is one of
the twenty PWN candidates within the central 20 pc \citep{mun2008}.

\parindent 0pt \hangindent 0pt \vskip 5pt
${\bf GCCR056}~-$ CXOGC J174536.9-290039 \citep{mun2009}, is a compact X-ray source having an
offset <0.2" from the radio source. In the deep X-ray catalog of \cite{zhu2018}, this source
is listed as  SS\#1767. The deep X-ray image shows that the bright compact X-ray source  appears to be at 
the end of a long (25") and slightly curved filament that extends to south  (see the top-right panel of Figure 3).
The compact radio source GCCR056  is embedded in extended radio source M which also has a filamentary 
component \citep{yusef1987}, but the long ($\sim$20") radio filament is oriented toward the  
northwest \citep{zmg2016}, so that the angle between the X-ray and radio filaments is about 120$^{\rm o}$. 

\parindent 0pt \hangindent 0pt \vskip 5pt
${\bf GCCR057}~-$ This source, located 3" SW of GCCR056,  coincides with a compact X-ray
source at the tip of  a linear feature that 
appears only in  the \cite{mun2008} catalog of extended X-ray sources; see the top-right 
panel of Figure 3. The linear X-ray source, G359.933-0.037, has  a power-law spectrum 
of $\Gamma=1.59$ and an X-ray luminosity of 3$\times10^{32}$ erg s$^{-1}$, which is 
one of the twenty suggested PWNs within the central 20 pc
\citep{mun2008}.

\parindent 0pt \hangindent 0pt \vskip 5pt
${\bf GCCR058}~-$ The radio source appears to be associated with a compact X-ray
source sunrounded by  extended emission source, G359.941-0.029 \citep{mun2008}.
The authors report 
a power-law spectrum of $\Gamma=0.44$ and an X-ray luminosity of 2$\times10^{32}$ erg s$^{-1}$.
The X-ray source is  one of the
twenty suggested PWNs within the central 20 pc \citep{mun2008}.

\parindent 0pt \hangindent 0pt \vskip 5pt
${\bf GCCR070}~-$ The X-ray counterpart CXOGC J174530.0-285942 \citep{mun2009}, 
which is offset by 0.7"  from the radio source and  is also found in 
\cite{zhu2018} as SS\#956. The deep Chandra image shows  the X-ray source having an amorphous halo 
with a size of $\sim4$" (see  middle-left panel of Figure 3).

\parindent 0pt \hangindent 0pt \vskip 5pt
${\bf GCCR072}~-$ The X-ray counterpart CXOGC J174528.8-285726  \citep{mun2009}, 
offset by <0.4" from the radio source, is also listed as 
SS\#852 in the X-ray catalog of  \cite{zhu2018}. The system is interpreted as an O-star in a 
colliding-wind binary (CWBs) or HMXB based on IR spectrocospy \citep{dew2013}.

\parindent 0pt \hangindent 0pt \vskip 5pt
${\bf GCCR089}~-$ The  X-ray counterpart CXOGC J174545.5-285828 \citep{mun2009}. 
The compact X-ray source is associated with an extended X-ray source that is interpreted as a
PWN \citep{par2005}. See the central panel of Figure 3. The radio emission from 
the PWN was described by \cite{zmg2013}. The compact source in 
both X-ray and radio   likely emanates from near the neutron star \citep{par2005,zmg2013}.

\parindent 0pt \hangindent 0pt \vskip 5pt
${\bf GCCR099}~-$ This radio source may be associated with a faint X-ray component
in the diffuse X-ray source,  G359.925-0.051 {\color{black} (see the top-middel panel in Figure 3).}   
It is interpreted as a PWN \citep{mun2008}. 
It is located $\sim$1" NE of GCCR047 (see Figure A1 and Figure 3).

\parindent 0pt \hangindent 0pt \vskip 5pt
${\bf GCCR106}~-$ This source is located 2" NW of GCCR057, and may be 
also associated with the candidate PWN, G359.933-0.037 \citep{mun2008}. 
See upper-right panel of Figure 3.

\parindent 0pt \hangindent 0pt \vskip 5pt
${\bf GCCR107}~-$ This  source  coincides (with an offset <0.5") 
with the X-ray source CXOGC J174536.1-285638 / SS\#1671 \citep{mun2009,zhu2018}. An 
investigation of the X-ray 
observations of the compact  X-ray source implies  an apparent 189$\pm$6 day periodicity 
present in the lightcurve \citep{mik2008}. The system is likely associated with an HMXB 
\citep{mik2008} or  with a colliding wind binary  based on its 
IR spectrum \citep{cla2009}. A spectral type of WN8-9h is suggested for the donor star \citep{mau2010}.

\parindent 0pt \hangindent 0pt \vskip 5pt
${\bf GCCR110}~-$ The  X-ray counterpart CXOGC J174532.7-285617 / SS\#1263  
\citep{mau2010,zhu2018} is offset by <0.7" from the compact radio source. 
Near infrared spectroscopy implies that the system is associated with a 
spectral type O4-6I star \cite{mau2010}. 
}

\section{Astrophysical implications}

\subsection{Spatial distribution \& extragalactic contribution}

The GCCR sources appear to be mainly distributed along the Galactic plane (Figure 1), indicating
that a significant fraction of the compact radio sources are located in the RBZ at the Galactic center. 
However, at the tens of $\mu$Jy level, the density of background extragalactic radio sources becomes 
noticeable. For example, the VLA deep observations at 5 GHz of the Great Observatories Origins Deep Survey - 
North (GOODS-N) ($\sigma_{\rm RMS}$=3.5 $\mu$Jy beam$^{-1}$, synthesized beam of 1.47"$\times$1.42")
and GOODS-S ($\sigma_{\rm RMS}$=3.0 $\mu$Jy beam$^{-1}$, with a beam of 0.98"$\times$0.45") 
fields found that these two fields contain 52 and 88 sources over areas of  109 and 190 arcmin$^2$, respectively 
\citep{gim2019}. The average source density in these two fields above a flux density of 15 $\mu$Jy is
therefore  $\sim$0.5 sources arcmin$^{-2}$.

From Table 2, a total  of 83 GCCR sources is found in a 45  arcmin$^2$ region within
the HPBW of the primary beam, excluding the area of 3 arcmin$^2$ covered by the Sgr A West and
Sgr A East HII regions. The density of GCCR sources above the 70 $\mu$Jy cutoff is therefore 
$\sim$1.8 sources arcmin$^{-2}$. If we use our GCCR cutoff of 70 $\mu$Jy to re-count the sources listed in the
GOODS-N and -S catalog \citep{gim2019}, the number of sources in the GOODS-N and -S surveys drops to 46, lowering
the source density to 0.15 sources arcmin$^{-2}$. Therefore, the density of GCCRs revealed by 
our search is an order of magnitude higher than that found in the GOODS-N and -S fields. 

We  note that the extragalactic source density of 0.15 sources arcmin$^{-2}$
at 5 GHz derived from the GOODS-N and -S fields is consistent with that of
$\sim$0.1 sources arcmin$^{-2}$ for extragalactic background sources above 100 $\mu$Jy 
at 3 GHz based on the derived source density by \cite{con2012}. Of course, the extragalactic 
background contribution is a function of the distance from the pointing center,
for a given flux-density cutoff, because it takes a stronger source to appear 
above the limit out at the edge of RBZ. In conclusion, we find that, at most, 
about 10\% of the GCCR sources are expected to be
associated with the extragalactic background population.

\subsection{Flux density distribution \& cataclysmic variables}
The Galactic center hosts a  large population of cataclysmic variables that 
are associated with hard X-ray sources \citep[e.g.][]{mun2004}. In a recent 
JVLA survey for radio emission from CVs, \cite{bar2017,bar2020} reported new 
detections of 33 magnetic CVs, or mCVs, with flux density in the range 
from 6 to 8031 $\mu$Jy at frequencies ranging between 4.5 and 22.1 GHz, 
increasing the number of radio sources associated with CVs to 40. 
The radio emission of the mCVs is circularly polarized \citep{bar2020} 
with relatively flat spectra \citep{bar2017}. Most of the radio CVs are
nearby, at distances  ranging from 88 pc to 2.24 kpc, spanning a 
radio luminosity range from 3$\times10^{24}$ erg s$^{-1}$
to 1.7$\times10^{27}$ erg s$^{-1}$.

\renewcommand{\thefigure}{4}
\begin{figure}[!h]
\centering
\includegraphics[angle=0,width=100mm]{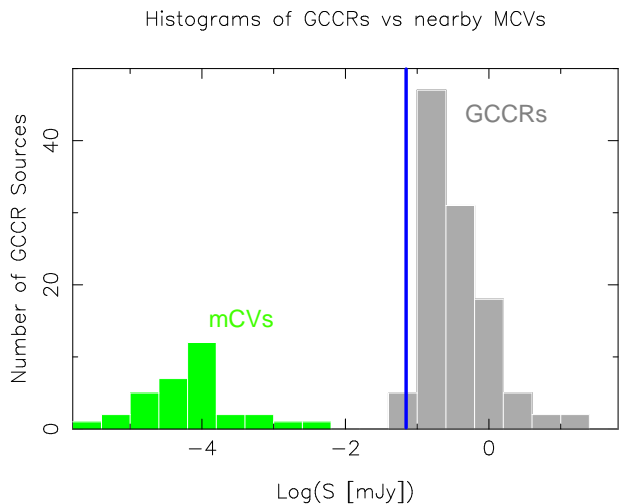}
\caption{Flux-density distributions of GCCRs (grey) versus the 33 nearby mCVs (green) 
detected with the VLA \citep{bar2020}. The flux densities of the mCVs in this histogram 
are from JVLA observations at frequencies between 4.5 and 22.1 GHz,
extrapolated to the GC at D=8 kpc assuming a flat spectrum with $\alpha=0$.
The blue vertical line marks the flux-density cutoff for GCCRs, 70 $\mu$Jy. 
}\end{figure}

To compare the flux-density distribution of the radio CVs with our GCCRs,
we scaled the radio flux density of CVs to the Galactic center by multiplying by 
${\rm (D/8~kpc)}^2$. Figure 4 shows a histogram of the radio source counts as 
a function of radio flux density in the logarithmic range between $-5.8$ and 1.8, 
corresponding to a range of flux density between 1.6 nJy (10$^{-6}$ mJy) and 63 mJy 
at a distance of 8 kpc; the logarithm of flux density, Log(S [mJy]), is binned
into $\Delta$ Log$(S [{\rm mJy}]) =0.4$ intervals starting from $-5.8$ (1.6 nJy).

The grey histogram in Figures 4 and {\color{black}5} shows a peak of 47 GCCRs between $-$1 to 
$-$0.6 in log (S [mJy]). No overlap in flux density is found between
the population of detected mCVs (green) and our reported sample of GCCRs (grey).

The source counts  below $-$1 (100$\mu$Jy) appear to be incomplete because only 
a small fraction of the GCCR candidates in the Log(S[mJy]) = -1.2 bin lie above our 
70 $\mu$Jy cutoff marked by the blue vertical line in Figure 4. In spite of the cutoff, 
the logarithmic flux-density distribution of the GCCR population shows 
a large dispersion, with an average of $\mu_{\rm Log(S[mJy])} = -0.44$ and an RMS of 
$\sigma_{\rm Log(S[mJy])} = 0.47$. The high-intensity tail of the distribution suggests
that the distribution of GCCRs may consist of multiple Gaussian or {\it normal} distributions 
of different source types. However, we can not rule out the possibility that the GCCR population 
shows an abnormal distribution of the compact radio sources.

\renewcommand{\thefigure}{5}
\begin{figure*}[!h]
\centering
\includegraphics[angle=0,width=190mm]{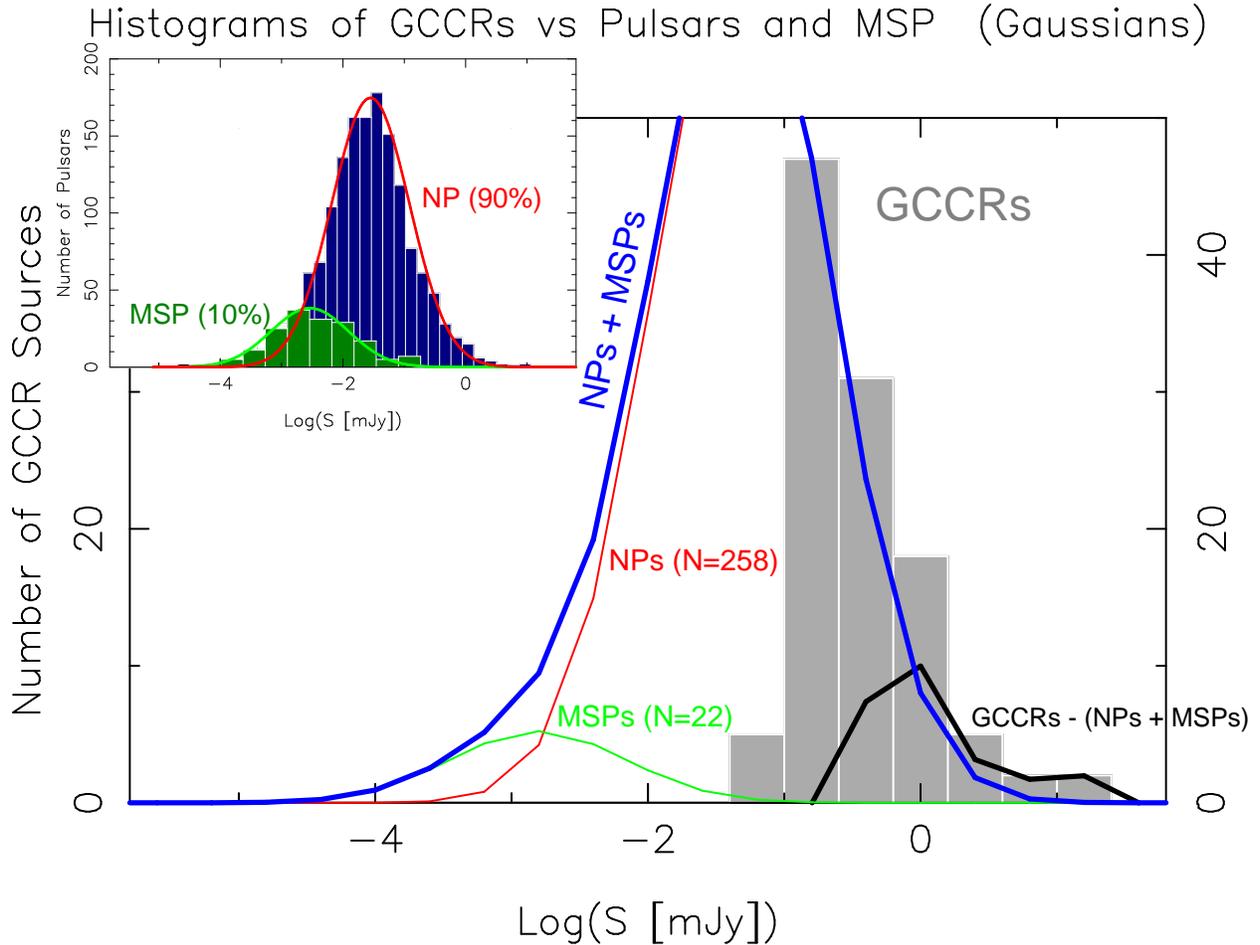}
\caption{The flux-density distribution of GCCRs (grey histogram) is fitted with a sample of 
258 NPs (red Gaussian curve) and 22 MSPs (green Gaussian curve), a 70\% of the Galactic disk 
(GD) NPs and MSPs sample  used in  \cite{kra1998}. The  blue curve is the  sum of 
the two Gaussians. The black curve denotes the difference between counts of the GCCR distribution
(grey histogram) and of the distrbution (the blue curve) that is 
extrapolated from the population of NPs and MSPs in the Galactic disk.
The 1.5 GHz flux densities of the NPs and MSPs from the samples of \cite{kra1998} have been 
scaled to the flux densities at 5.5 GHz assuming $\alpha=-1.7$. The \cite{kra1998} sample 
has also been scaled to the Galactic center distance by scaling their flux densities by 
${\rm [1~kpc/8~kpc]}^2$. {\color{black} The top-left inset shows the distributions of
1503 NPs (dark blue histogram) and 169 MSPs (dark green histogram) 
based on a large sample observed at 1.4 GHz \citep{man2005}.  
The 1.4 GHz flux densities have also been scaled to the flux densities at 5.5 GHz in the Galactic
center assuming $\alpha=-1.7$ and the GC distance of 8 kpc. The red (NPs) and green (MSPs) curves
show the fitted Gaussian distributions with $\mu_{\rm NP}=-1.3$ and $\mu_{\rm MSP}=-2.5$ as well as a common 
standard deviation of $\sigma=0.64$. 
}
}\end{figure*}

\subsection{Normal pulsars and MSPs at the Galactic center}
We consider here the possibility that some of the GCCRs could be pulsars.  
While the present formation rate of massive stars in the GC is large enough 
to give rise to the expectation that pulsars would be abundant in the GC, 
very few are known, {\color{black} presumably because  the foreground 
scatter broadening toward the GC} \citep[{e.g.,}][]{spi2014} leads in 
most cases to a sufficiently large pulse broadening that the pulses become 
indistinguishable. However, with sufficient sensitivity, pulsars can be 
detected as point-like continuum radio sources or as pulsar wind nebulae.

A comparison of luminosities and spectral indices between samples of normal 
pulsars (NP) and millisecond pulsars (MSP) has been conducted  by \cite{kra1998} 
based on 31 MSPs and 369 NPs distributed in the Galactic disk (GD) 
\citep[Also see][]{lor1995,tay1993}. They showed that NPs and MSPs have similar spectra,  
with spectral indices of $\alpha=-1.6\pm0.04$ and $\alpha=-1.8\pm0.1$, respectively. 
In addition, the MSPs are an order of magnitude less 
luminous than NPs. A mean value of Log(Sd$^2$ [mJy kpc$^2$]), $\mu_{\rm MSP}=0.5\pm$0.2 at $\sim$1.5 GHz, 
is derived for MSPs as compared to $\mu_{\rm NP}=1.5\pm$0.04 for the NPs \citep{kra1998}. 
{\color{black} As noted by \cite{kra1998}, the statistics may be subject to a bias owing to the fact that
most NPs were discovered in surveys at higher frequencies (that correspondingly selected flatter
spectrum and more luminous pulsars), and that most MSPs were discovered
at low frequencies and were therefore relatively nearby subject to
limitations in dispersion removal.} 
To avoid possible statistical bias caused by the difference in the {\color{black} observed} 
luminosities of MSPs and NPs, \cite{kra1998} investigated a statistically complete sample of nearby MSPs and NPs. 
{\color{black} They} demonstrated that the discrepancy of the mean values between MSPs and NPs
becomes small in the case of restricting to a nearby population within a distance of $1.5$ kpc. They 
find that the mean values of $\mu_{\rm MSP}$ and $\mu_{\rm NP}$  are $0.0\pm$0.1 and $0.57\pm$0.09 at 
$\sim$1.5 GHz, respectively, in a nearby population of 18 MSPs and 55 NPs after removing the apparent biases.

However, the nearby sample excludes the high-luminosity NPs and MSPs that may make a significant contribution 
to the GCCR population. We would need a large number of pulsars ($\sim3000$) to fit the upper 
tail of the GCCR distribution 
if we scaled the nearby sample of \cite{kra1998} to the Galactic center. 
The GD population appears to be more relevant to the distribution of GCCRs. Using the spectral index  
$\alpha=-1.7$ for both MSPs and NPs and a distance of 8 kpc for the Galactic center, we extrapolated 
the mean values of $\mu_{\rm MSP}$ and $\mu_{\rm NP}$ at 1.5 GHz of the GD population of MSPs and NPs 
to the corresponding values at 5.5 GHz for the GC population, giving $\mu_{\rm MSP}=-2.3$ (5$\mu$Jy) 
and $\mu_{\rm NP}=-$1.3 (50$\mu$Jy) at 5.5 GHz. We then compare the extrapolated GD populations of 
{\color{black} 369} NPs and {\color{black} 31} MSPs to the distribution of GCCRs in Figure 5, 
by approximating the pulsar distributions 
as Gaussian with a common standard deviation of $\sigma=0.64$; the value of $\sigma$ was estimated from 
the FWHM of the NPs' distribution in Log(S$\rm d^2 [{\rm mJy~kpc^2}])$ (Fig. 2 of Kramer et al. 1998).
Therefore, on the tentative assumption that all the GCCRs in the 100-250 $\mu$Jy bin are pulsars, except 
for the 10\% of them expected to be extragalactic sources,  a total of 22 MSPs and 258 NPs would be
needed to account for the GCCR distribution. Namely, 70\% of  the GD population, used in
the \cite{kra1998} analysis, would be required to match the 47 GCCRs detected in the Log$(S [{\rm mJy}])=-1$ 
bin covering the flux density range 100-250 $\mu$Jy. 

{\color{black} We further inspected and verified the statistics of \cite{kra1998}  with a 
large sample of 1672 pulsars observed at 1.4 GHz \citep{man2005}, 90\% of which is NPs 
(spin period $P>30$ millisecond) and 10\% is MSPs ($P \le 30$ millisecond); see the inset of 
Figure 5. Scaling to the flux densities at 5.5 GHz at the Galactic center distance 
(D$=8$ kpc), and  assuming $\alpha=-1.7$, we derive the mean ($\mu_{\rm NP}=-1.3$) 
and standard deviation ($\sigma=0.64$) of the logarithmic flux density from the 1503 NPs, that are 
in good agreement with the corresponding parameters derived from the GD sample of \cite{kra1998}. 
The mean logarithmic flux density of MSPs ($\mu_{\rm MSP}=-2.5$) derived from the 169 MSPs 
is slightly less than the value ($\mu_{\rm MSP}=-2.3$) of the GD sample, indicating that a difference 
in the mean flux density  between NPs and MSPs in the large sample is
insignificantly greater than that of the GD sample.}    

Therefore, the analysis here is consistent with the possibility that up to 80\% of the detected GCCRs could 
be  NPs if the RBZ hosts a total of 280 NPs and MSPs with 
a distribution in radio luminosity similar to the GD distribution of NPs and MSPs. However, 
the MSP population essentially makes no contribution to
the upper tail of GCCR sources detected in this paper. 
Of course, the possible number of NPs among the GCCRs given above is an upper limit, 
as other classes of sources can also contribute to the GCCR population, notably the X-ray binaries that we discuss below.   
A first filter for constraining the NP population among the GCCRs could be based on spectral index measurements, 
given the typically steep spectra of NPs ({\color{black}$\alpha~\sim~-1.7$}). 
 We also note that NPs are usually not strongly variable on time scales of 6 years or shorter 
(Paul Demorest, personal communication) and only 25\% of the GCCRs are non-variable, 
so it appears that NPs are, at most, a minor fraction of the GCCRs.  
Of course, firmly identifying pulsars requires detection of their pulsed emission.
To date, PSR J1745-2900 is 
the only confirmed pulsar within the RBZ.  PSR J1745-2900 was first identified as an X-ray source 
by the Swift observatory during a flare \citep{ken2013}; and  pulsed emission with a period 
of 3.76 s  was revealed in follow-up observations by the  NuSTAR observatory \citep{mor2013}.
We note that the analysis in this section does not cover the compact radio sources located within Sgr A West, 
and the Sgr A East HII complex. Located $\sim$3" away from Sgr A*, PSR J1745-2900 is not listed in Table 2,
our GCCR catalog. The discovery of PSR J1745-2900, the GC magnetar,             
raises the possibility that it might be possible to detect 
pulsed emission from some of the GCCRs.           

\subsection{X-ray binaries}

By comparing the GCCRs in our 5.5-GHz image with published catalogs of X-ray sources based 
on observations with the Chandra X-ray observatory, and with the Chandra X-ray image from 
Zhu et al. (2018), we find about 42 possible X-ray counterparts to the GCCRs (Figure 3). The GCCRs 
identified with X-ray counterparts could be close binary systems in which a compact stellar remnant 
accretes mass from its companion.

X-ray binaries can be divided into two major spectral states based on the hardness
of their X-ray spectra: soft and hard states. The soft state is dominated by thermal 
emission from an accretion disk, while the hard state is dominated by the emission 
from the corona \citep{cor2011}. The radio emission in the hard state is usually 
characterized by a flat or slightly inverted spectrum with a spectral index of
$\alpha\sim0$, which can be interpreted as self-absorbed synchrotron emission
from a compact jet, similar to those found in extragalactic nuclei \citep[{\rm e.g.,}][]{bla2019}.
During the soft state, the compact jets are likely to be quenched \citep[{\rm e.g.,}][]{fen1999,cor2011}.
The presence of a strong correlation between radio and X-ray emission during the hard state
has been investigated with observations of several X-ray binaries 
\citep[{\rm e.g.,}][]{cor2000,mig2006,cor2011,tud2017,gal2018,qia2019}, showing
a power-law relationship ($L_{\rm R}\propto\/L_{\rm X}^\beta$) between the luminosities 
of X-ray ($L_{\rm X}$) and radio ($L_{\rm R}$).

For black hole X-ray binaries (BHXB) \citep{fen2009}, the standard value for the power-law index,
 $\beta\sim0.6$ \citep{cor2003,cor2008,gal2003,xue2007,cor2011} is thought to be related to the inner region of the accretion 
system where a hot and inefficient accretion flow (i.e., an advection-dominated accretion flow, or ADAF) might 
be present \citep{nar1994,nar1997,abr2013}. The ADAF model appears to reasonably account for sources in the hard state
while the radio emission is optically thick and is correlated with X-ray emission. On the other hand, a steady, 
powerful, relatively low bulk velocity or bulk Lorentz factor $\Gamma<2$ jet is always present in the hard X-ray 
state \citep{fen2009}. The observed jets imply  a combination of radiatively inefficient flows 
with the simultaneous presence of MHD winds
or outflows. That is, advection-dominated inflow-outflow solutions, 
or ADIOS \citep{bla1999}, may work for the BHXBs.

Similar power-law correlations between $L_{\rm X}$ and $L_{\rm R}$ are also shown by neutron-star
X-ray binaries (NSXB), but BHXBs are more radio loud by a factor of 20-30 \citep{gal2018,kyl2012}. 
In a study of disk-jet coupling in low-luminosity accreting NSs in LMXBs, \cite{tud2017} show 
$L_{\rm R}\propto\/L_{\rm X}^\beta$ relations characteristic of three different types of NSs as compared to
the standard relation $\beta\sim0.6$ for BHXB. Transitional millisecond pulsars (tMSP) show $\beta\sim0.6$, 
the same as that for BHXBs but with an order of magnitude less luminosity at 5-GHz than BHXBs.
Non-pulsing NSs correspond to $\beta\sim0.7$ while hard-state NSs have $\beta\sim1.4$ \citep{tud2017}.
It is worth mentioning that the data used in their analysis span six orders of magnitude in 5-GHz radio 
luminosity ($L_{\rm 5~GHz}:  10^{25-31}$ erg s$^{-1}$) and nine orders of magnitude in the 1-10 keV X-ray 
luminosity ($L_{\rm X}: 10^{30-39}$ erg s$^{-1}$.) The BHXBs are mainly distributed in the range of 
$L_{\rm 5~GHz}$:  (10$^{28-31}$ erg s$^{-1}$) along the power-law correlation curve 
($L_{\rm 5~GHz}\propto\/L_{\rm X}^{0.6}$), while the NSXBs are clustered in a domain around  
10$^{27-29}$ erg s$^{-1}$ in $L_{\rm 5~GHz}$ and a few times 10$^{36-37}$ erg s$^{-1}$ in $L_{\rm X}$. 
For X-ray binaries having luminosities in the range of 10$^{36-37}$ erg s$^{-1}$, the BHXBs appear 
to be distinguishable from the counterpart NSXBs based on their much higher radio luminosities.

\renewcommand{\thefigure}{6}
\begin{figure}[!h]
\centering
\includegraphics[angle=-90,width=100mm]{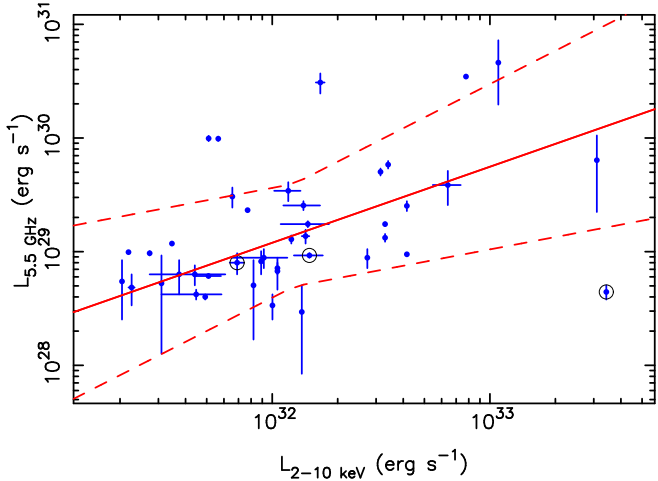}
\caption{Plot showing the cross-correlation between X-ray and radio luminosities for 
the 42 GCCRs having X-ray counterparts.
The radio luminosities are derived from the flux densities at 5.5 GHz given in this paper 
and the X-ray luminosities are derived from the X-ray photon fluxes in the 2-8 keV X-ray band reported in the
X-ray catalogs \citep{mun2008,mun2009,zhu2018}.  The black circles mark the three candidate HMXBs (Section 4.2).
The red solid line is the result from a least-squares fit to the logarithmic X-ray and radio luminosities, 
implying a power-law relation $L_{\rm 5.5~GHz}  \propto L_{\rm 2-10~keV}^\beta$. {\color{black}The red dashed lines outline 
the range in which the true regression line lies at a confidence level of 
95\%, derived with Scheff\'e's method, 
(https://en.wikipedia.org/wiki/Confidence\_and\_prediction\_bands
   and
https://www.itl.nist.gov/div898/handbook/prc/section4/prc472.htm).}
}
\end{figure}
     
We carried out a regression analysis for the cross-correlation between logarithmic radio and X-ray luminosities for the
42 GCCRs with X-ray counterparts. The radio luminosities are derived from the flux densities $S_\nu$ given in 
Table 2 using the form $L_\nu=4\pi D^2\nu S_\nu$ and $\nu=5.5$ GHz. The X-ray luminosities are derived from the 
photon flux values $S_{\rm 2-8~keV}$, listed in Table 2, provided in the Chandra X-ray catalogs  
\citep{zhu2018,mun2009,mun2008} using the form $L_{\rm  2-8~keV}= 4\pi D^2S_{\rm 2-8~keV} f_{\rm 2-10~keV}$, 
where $f_{\rm 2-10~keV}$ is a photon flux-to-energy conversion factor. We adopted $f_{\rm 2-10~keV}= 2.38 \times 10^{-8}$ 
erg photon$^{-1}$ \citep{zhu2018} to compute the 2-10 keV unabsorbed energy flux.
Figure 6 shows a plot of $L_{\rm 5.5~GHz}~vs.~L_{\rm 2-10~keV}$ for the 42 GCCRs with X-ray counterparts.
We performed a least-squares regression analysis assuming a linear relationship between the logarithmic 
radio and X-ray luminosities, $\log \big(L_{\rm 5.5~GHz} {[\rm erg~s^{-1}]}\big) = 
\alpha +\beta \log \big(L_{\rm 2-10~keV} {[\rm erg~s^{-1}]}\big)$.
We find that $\alpha=7.7\pm0.45$ and $\beta=0.67\pm0.02$ with a correlation coefficient R=0.72 and
a probability of no correlation, P~<~0.01\%. The $\beta$ value derived for the GCCRs appears 
to be consistent with the power-law relationships  that are found for  BHs, tMSPs, and non-pulsating NSs  
in the LMXB sample used in the analysis of \cite{tud2017}.

We also note that the 5.5-GHz radio luminosities of the GCCRs with X-ray counterparts
are in the luminosity range of 10$^{28-31}$ erg s$^{-1}$,  consistent with the range of
5-GHz radio luminosities of the BHXBs used in the analysis of \citep{tud2017}.
However, about twenty GCCRs with X-ray counterparts having 5.5-GHz radio luminosities 
below $1\times10^{29}$ erg s$^{-1}$  could be explained as NSXBs.
The five radio morphology types (given in column 12 of Table 2, Appendix A) of the GCCRs are also consistent
with the possibility that the compact radio cores are produced from either BHXBs or NSXBs.
If the compact cores of the GCCRs are associated with jet flows from the inner region of 
accretion disks or from the corona of compact objects, their radio spectra are expected to be flat \citep{cor2011}. 
In addition, a fraction of GCCRs associated with pulsars and MSPs discussed in Section 5.3 
may belong to the catagory NSXBs, if they are binaries emitting X-rays. However, some of the GCCRs 
with X-ray counterparts listed in Table 3 may just be associated with PWNs powered by a single neutron star.
Further study of the GCCRs with {\it coordinated  radio and X-ray  observations} will help to distinguish
between BHs and NSs for the compact objects associated with the GCCRs.

Finally, we note that the recent detection of  a 91$\pm10$ $\mu$Jy source at 5.5 GHz at the position 
of the Galactic center transient  caught during its flare in 1990 (GCT1990)  with a flux density then 
of $\sim$1 Jy at 1.5 GHz \citep{zha1992,zmg2020}, implying $L_{\rm 5.5~GHz}=2\times10^{32}$ 
erg s$^{-1}$ during the 1990 flare. This GCCR source is located within the Sgr A West region, 
which does not match the selection criteria used to compile Table 2; so the GCT1990 is not included in the 
above analysis. If this radio source is a remnant or the impact site of the compact jet of 
GCT1990 that has been quenched as the source transitioned from a hard state to the soft state, 
then the quenching factor\footnote{The quenching factor is defined as a ratio of the peak 
value of radio flux density during an outburst to the lowest value in the outburst light curve 
\citep{cor2011}.} of the GCT1990 is $\sim$5000, an order of magnitude greater than  that of 
H1743-322 \citep{cor2011}. The high radio luminosity during the outburst of 1990 is consistent
with the hypothesis of a BHXB for the GCT1990 \citep{zha1992}, although the possibility of a 
NSXB cannot be completely ruled out.

\section{Conclusion}

We imaged the RBZ with wideband continuum data taken at 5.5-GHz with the VLA in its
A array at three epochs: 2019-9-8, 2014-5-17 and 2014-5-26. A total of 110 GCCRs 
has been detected at an angular resolution of 0.4" outside Sgr A West and the 
complex of Sgr A East HII regions. The 10$\sigma$ cutoff in flux density used in the GCCR survey
is 70$\mu$Jy. Five types of sources are classified according to their morphology: 
1) an unresolved source; 
2) a compact source   with a size determined from 2D Gaussian fitting; 
3) a compact source  associated with a linear feature;
4) a compact source with a radio tail; and 
5) a double compact source.

In general, the GCCRs are distributed along the Galactic plane and about ten percent of them are
expected to be extragalactic background sources. 
The mean value of logarithmic flux density at 5.5 GHz, $\mu_{\rm Log(S[mJy])}=-0.44$ with a standard deviation
of $\sigma_{\rm Log(S[mJy])}=0.47$, 
suggests that the GCCRs are at least three orders of magnitude more luminous than the radio sources
powered by magnetic cataclysmic variables, i.e., close binaries containing a WD.  On the other hand, when
compared to the Galactic disk (GD) population of normal pulsars (NPs), 
a majority (80\%) of the GCCRs appears to fall within the high flux-density tail of the pulsar distribution, as 
extrapolated from a sample of NPs in the Galactic disk. However,  MSPs extrapolated 
from the GD population are too weak to have contributed significantly to the GCCR population
that have been detected.  

We also cross-correlated the GCCRs with X-ray sources in Chandra X-ray catalogs and
found that 42 GCCRs have candidate X-ray counterparts. In addition, our regression analysis shows
that the logarithmic X-ray ($L_{\rm 2-10~keV} [{\rm erg~s^{-1}}]$) and  
radio ($L_{\rm 5.5~GHz} [{\rm erg~s^{-1}}]$) luminosities are linearly correlated,
with a correlation coefficient of 0.72.
The radio luminosities and radio morphologies, along with the compactness of the sources, suggest 
that the radio emission from the GCCRs having X-ray counterparts is consistent with compact radio 
jets launched from X-ray binary systems associated with either a black hole or a neutron star.
Some of them are associated with PWNs. Among the GCCRs with candidate X-ray counterparts, 
the lower luminosity ones could include some NSXBs while those with a higher luminosity are candidate BHXBs.

\acknowledgments
{{\color{black}We are grateful to the anonymous referee and to the editors for 
providing their valuable comments and suggestions.} The Very Large
Array (VLA) is operated by the National Radio Astronomy Observatory
(NRAO). The NRAO is a facility of the National Science Foundation
operated under cooperative agreement by Associated Universities, Inc.
The research has made use of NASA's Astrophysics Data System. {\color{black}This
research has also made use 
of the VizieR catalogue access tool, CDS, Strasbourg, France (DOI: 10.26093/cds/vizier).} 
}
\vskip 40pt
\appendix
\section{the Galactic center compact radio sources}

We catalog the newly detected 110 GCCRs from the RBZ, covering the central  180 arcmin$^2$ area.
The radio flux densities of individual GCCRs are determined from the three epochs observations
on 2019-9-8, 2014-5-26 and 2014-5-17. 
The 110 compact sources listed in Table 2 are divided
into two groups: (1) Variables or transients ($N=82$) if $\Delta S/{\sigma}\ge4$, 
where 
\begin{equation}
\Delta S=S_{\rm max}- S_{\rm min}
\end{equation}
is the range of variation in flux density S
and 
\begin{equation}
\sigma=\big(\sum_{i=1}^{i=3} \frac{1}{\sigma_i^2}\big)^{-1/2}
\end{equation} 
is an uncertainty in the  average  flux density,  S.

{\parindent 0pt
(2) Non-variables ($N=28$) if $\Delta S/{\sigma}<4$.
}

\subsection{A catalog of the GCCRs}
Table 2 lists the radio properties of the 110 GCCRs along with their X-ray identifications.

Column 1 is the source ID for the
Galactic center compact radio sources (GCCR) that are identified from the three epoch's
VLA 5.5-GHz images; these images  are 
produced by filtering out
the short-spacing uv data
and applying the correction for PB attenuation.
The three dirty images  were cleaned with the MS-MSF algorithm \citep{rau11}
and the images with the cleaned components were finally convolved with a common beam of
FWHM 0.50"$\times$0.24" ($-0.05^{\rm o}$). Thus, the 
intensities of a source
observed at the three epochs are not biased by different sizes of their original 
synthesized beams that are listed in column 9 of Table 1.
A comparison of the source intensities can be carried out for the three epochs.
Figure A1 shows the 5.5-GHz, high-resolution contour images for each of the GCCR
sources at the three epochs displayed in the same column: 
panel (a) for 2019-9-8, (b) for 2014-5-26, and (c) for 2014-5-17.
The images were made with MS-MFS algorithm \citep{rau11} averaging every two
channels with a resultant channel bandwidth of 4 MHz.

Column 2 gives the  equatorial coordinates of the sources at the epoch of J2000.
The uncertainty in position presuambly dominated by thermal noise is
$\sigma_\theta=0.5\theta_{\rm beam}{\rm SNR}^{-1}$, where $\theta_{\rm beam}$ is
a FWHM of telescope beam and SNR is a ratio of signal-to-noise.
Given a synthesized beam  of  $\theta_{\rm beam}\sim$0.6, elongated nearly in N-S,
and minimum SNR of 10, the positional uncertainties in RA and Dec are
$\sigma_\alpha \lesssim 0.002^s$  and $\sigma_\delta \lesssim  0.03$", respectively.
However, a source located far from  the phase center of the interferometer array
is subject to a bandwidth smearing (BWS) effect$^{5}$. 
Thus, the clean beam is smeared by a Gaussian in the radial direction, with a
 FWHM proptional to $\frac{\delta\nu}{\nu_0}r_\theta$. For a source located at the edge of the
field  $r_\theta\sim$450", the quantity $\frac{\delta\nu}{\nu_0}r_\theta\sim$0.32" represents 
the largest angular size caused by the  BWS effect corresponding to the ratio of channel width to 
band-center frequency $\frac{\delta\nu}{\nu_0}=7\times10^{-4}$.
Convolving $\theta_{\rm beam}$ with  the FWHM of the BWS effect, the resulting beam will increase
by a factor $<1.13$, depending on $\frac{\delta \nu}{\nu_0}r_\theta$.

Column 3 lists the angular distance $r_\theta$ of a GCCR source with respect to the phase center
($\alpha_{\rm p}$, $\delta_{\rm p}$) for given GCCR source RA and Dec ($\alpha$, $\delta$)
based on  the following equation:
\begin{equation}
\cos({r_\theta}) = \cos(\delta)\cos(\delta_{\rm p})+\sin(\delta)\sin(\delta_{\rm p})
\cos(\alpha-\alpha_{\rm p}).
\end{equation}

Columns 4 and 5 give the angular offsets in RA and Dec with respect to Sgr A*.

Column 6 lists
the PB correction factor $\mathscr{F}_{\rm PB}= A(x)^{-1}$.

Column 7 gives $\sigma\mathscr{F}_{\rm PB}^{-1}=\sigma_A\/A(x)^{-1}$, corresponding to
the fractional uncertainty of the PB correction. The uncertainty $\sigma_A$ of $A(x)$ is computed
with Eq(2).

As a consequence of the BWS effect of stretching the synthesized beam, 
the apparent peak intensity of a source decreases, while the source
flux density remains invariant. The source intensity is reduced by a factor of
$\sqrt{1+\beta^2}$, where 
\begin{equation}
\beta=\frac{\delta\nu}{\nu_0}\frac{r_\theta}{\theta_{\rm beam}}
\end{equation} 
for a FWHM synthesized beam $\theta_{\rm beam}$ \citep{bri99}.
For a 2D Gaussian source, the flux density $S$ of a source is a linear function of the apparent peak intensity
$S_{\rm p}$ and angular size $\Theta^{\rm }_{\rm FWHM}$,
\begin{equation} 
S=\frac{\pi S_{\rm p}}{4\ln(2)}  {\rm \Theta^{\rm }_{\rm FWHM}}. 
\end{equation}
The apparent angular size $\Theta^{\rm }_{\rm FWHM}$ is a resultant of  
the source intrinsic  size ($\theta_{\rm maj}\times\theta_{\rm min}$) convolved with a telescope beam.
In principle, the smearing effect reduces  $S_{\rm p}$  and enlarges
$\Theta^{\rm s}_{\rm FWHM}$ but does not change $S$. The AIPS task 
{\sf JMFIT} provides an option for correcting the BWS effect
while fitting a 2D Gaussian function to a compact source. The flux densities S
along with the uncertainties $\sigma$ due to the RMS noise are reported in columns 8 to 10
corresponding to the measurements at the epochs 2019-09-08, 2014-05-26 and 2014-05-17.

Column 11 provides the RMS $\sigma_{\rm map}$ in the regions near the sources 
that are plotted in contours, see Figure A1.

Column 12 gives classifications of the GCCR sources. 
Five types of sources are classified according to their morphology:
{\sf u-core} stands for unresolved compact source; {\sf c-core} is for a compact source with a size determined from
2D Gaussian fitting;
{\sf l-core} is for a compact source associated with a linear feature;
{\sf t-core} is for a compact source having a tail, and {\sf d-core} is for double compact source. The results
derived from 2D Gaussian fitting for intrinsic sizes $\theta_{\rm maj}$ and $\theta_{\rm min}$,
as well as position angle PA are given in the notes for corresponding individual sources for all the GCCR types other
than {\sf u-core}.

Column 13 provides a brief note for the X-ray identifications. The code "y" stands for the GCCR sources that are
identified with X-ray counterparts with a positional offset between X-ray and radio less than 1 arcsec 
($\Delta\theta_{X-R}<$1") or located in the inner region of an X-ray halo; the code "y?" means that a possible X-ray counterpart is present near the 
GCCRs or $\Delta\theta_{X-R}=$1" to 2" for the offsets between the GCCRs and X-ray candidates; 
the letter "n" means that no X-ray counterparts have been identified for the GCCRs with
$\Delta\theta_{X-R}>$2".  The  procedure to identify  X-ray counterparts for the GCCR sources was based 
on cross-examinations between the Chandra X-ray and VLA 5.5-GHz images 
in addition to searching the online catalogs of the X-ray sources at the Galactic center 
\citep{mun2008,mun2009,zhu2018} for the  GCCRs' X-ray counterparts as described in section 4.

\include{appendix}

\end{document}

%% file: appendix.tex
\renewcommand{\thefigure}{A1}
\begin{figure*}[tbh]
\centering
\includegraphics[angle=0,width=85mm]{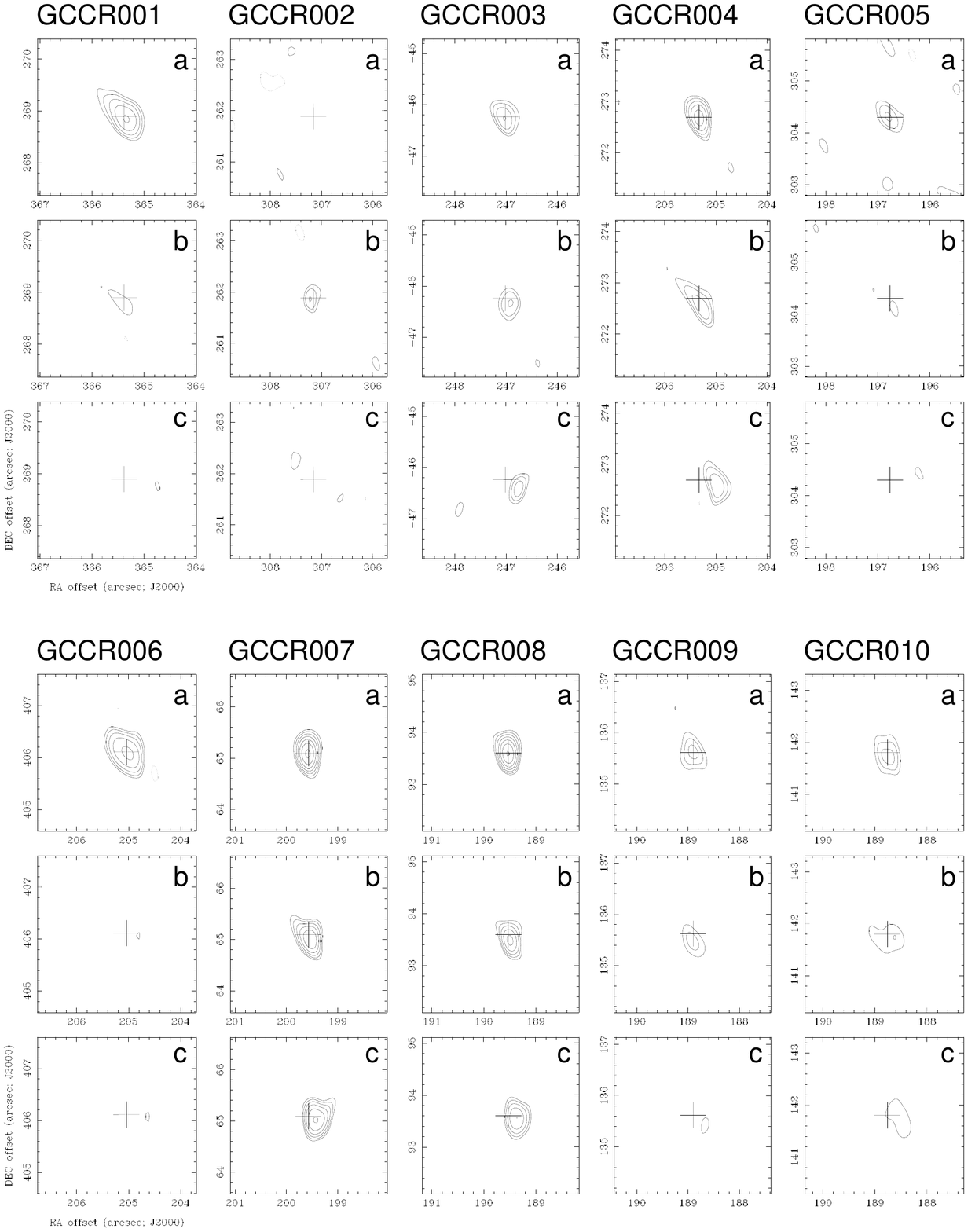}
\includegraphics[angle=0,width=85mm]{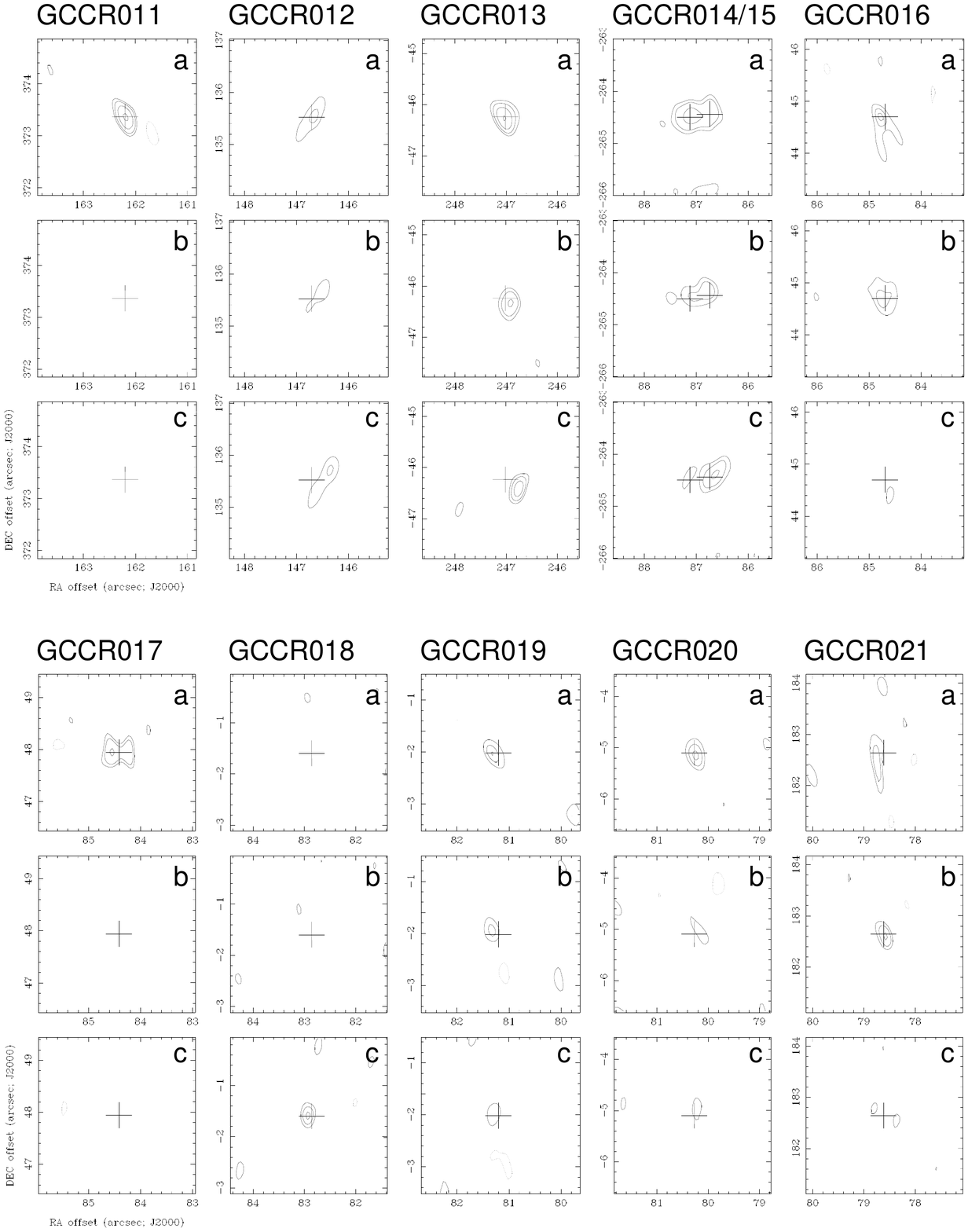} \\
\includegraphics[angle=0,width=85mm]{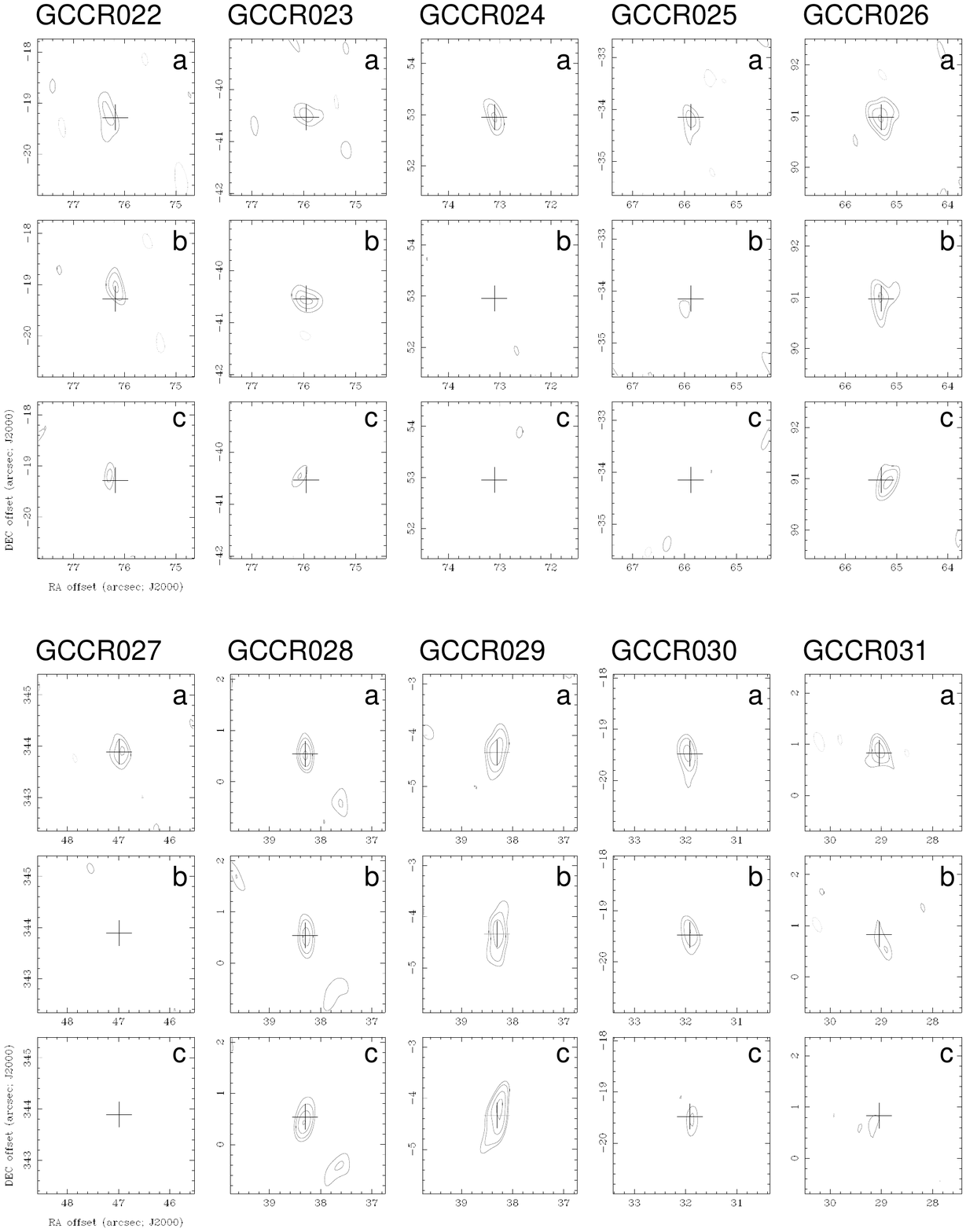}
\includegraphics[angle=0,width=85mm]{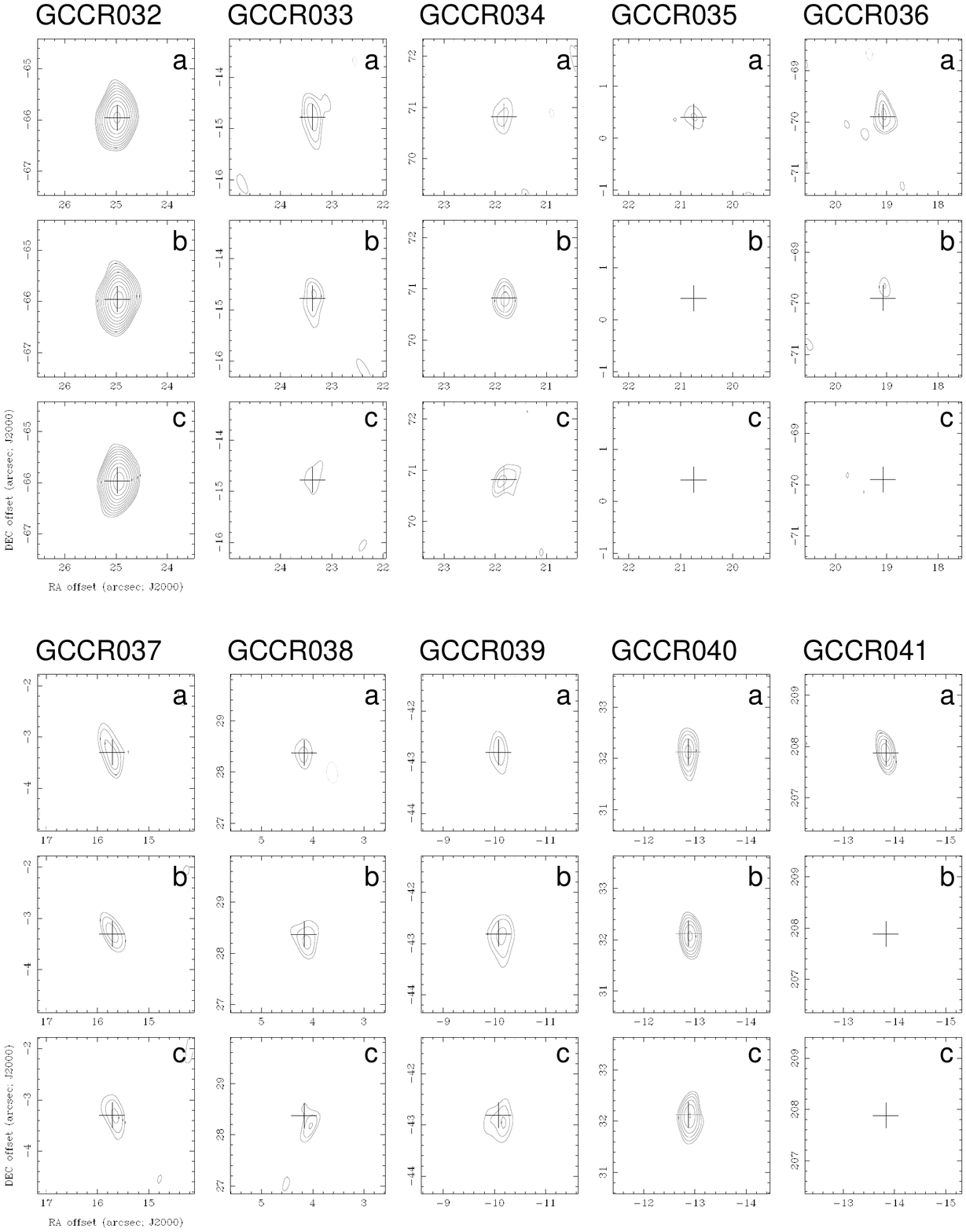} \\
\caption{
Contour plots for individual GCCR sources at epochs 2019-9-8 ({\bf a}),
2014-5-26 ({\bf b}), and 2014-5-17 ({\bf c}). Each of the sources is
arranged in the same column labled with its GCCR ID number at top.
Contours are $\sigma_{\rm map} \times$ ($-5$, 5$\times\sqrt{2^n}$), where $n=-1$, 0, 1, 2, 3, 4, 5 ... until reaching
the intensity peak.
The rms noises $\sigma_{\rm map}$ are listed in column 11 in Table 2 for corresponding GCCR sources.
For the panels of Sgr A*, $\sigma_{\rm map}=10$ mJy beam$^{-1}$, a hundred times greater
than the local rms noise near the source. The coordinate labels are the angular offsets
from the phase center of the date or the pointing center of the VLA observations; the coordinates
of the phase center are given in the footnote$^3$ in text.
}
\end{figure*}

\renewcommand{\thefigure}{A1}
\begin{figure*}[tbh]
\centering
\includegraphics[angle=0,width=85mm]{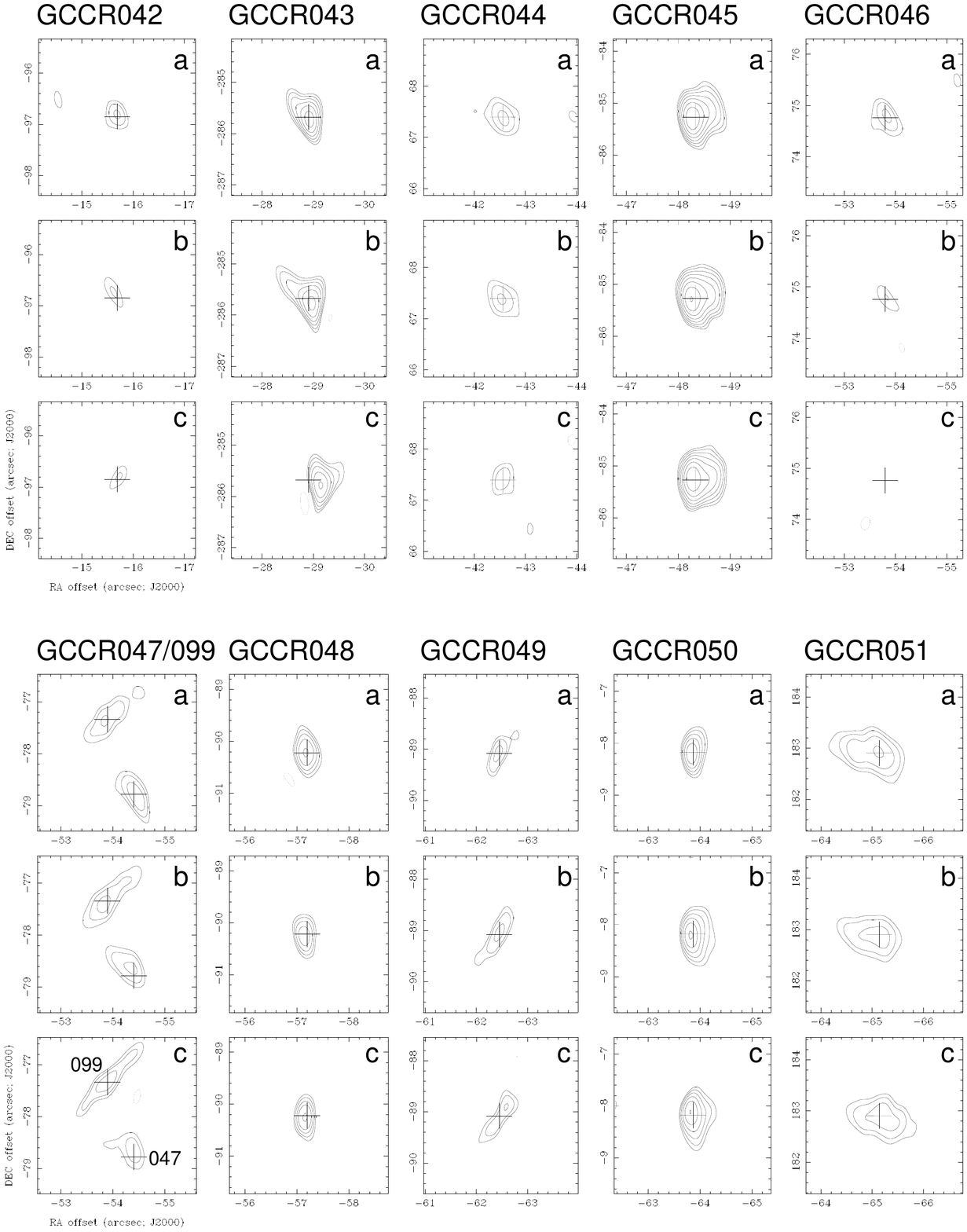}
\includegraphics[angle=0,width=85mm]{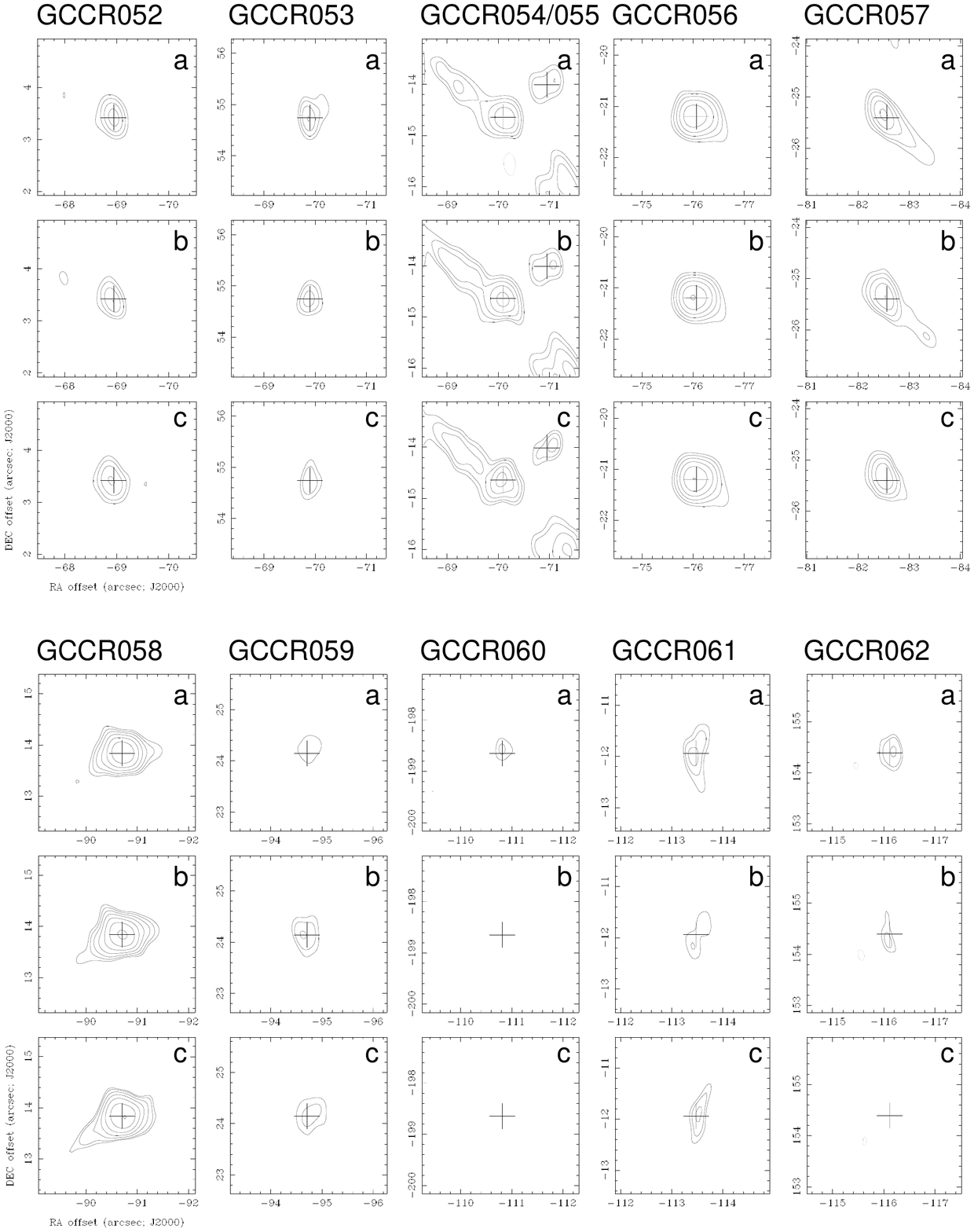} \\
\includegraphics[angle=0,width=85mm]{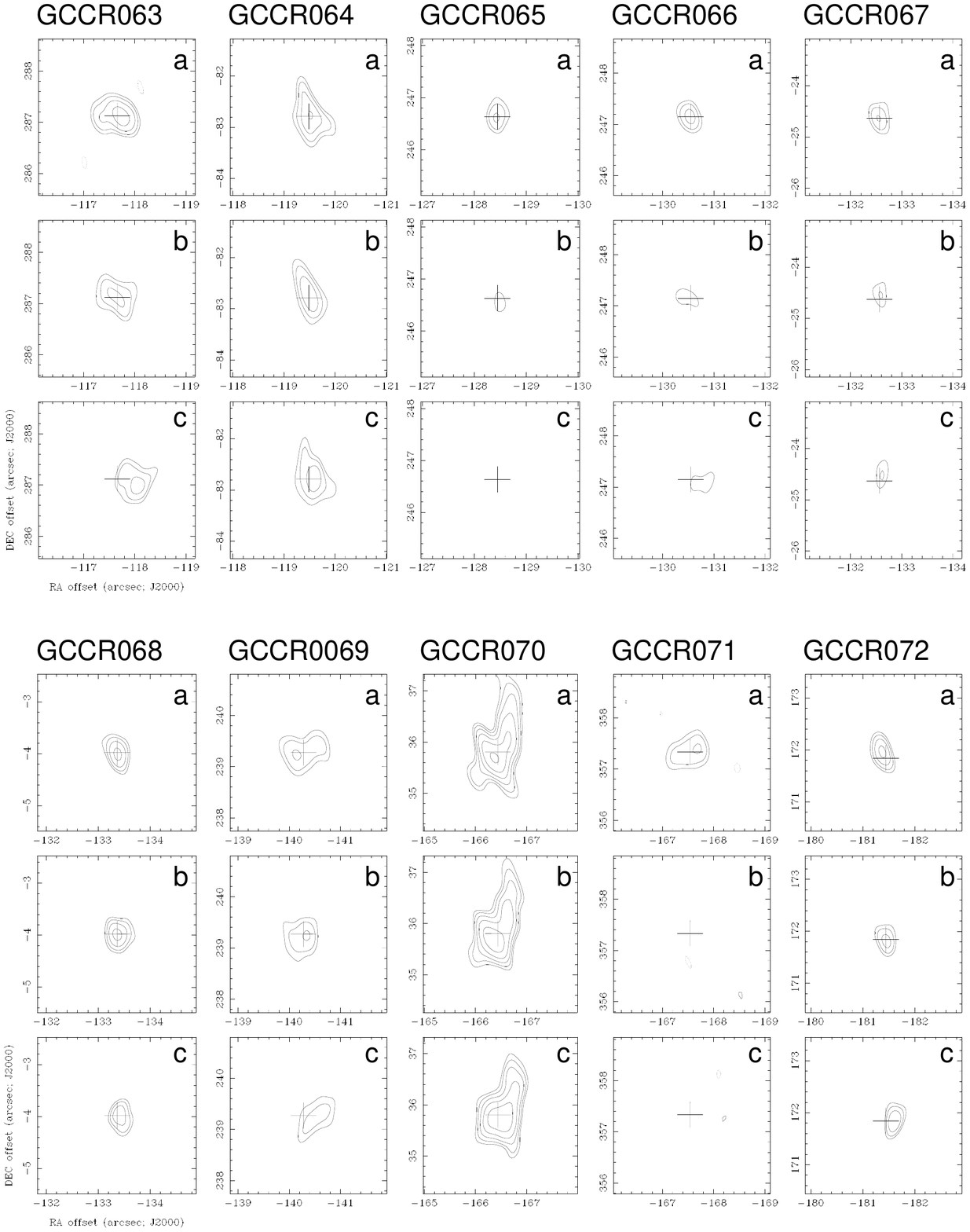}
\includegraphics[angle=0,width=85mm]{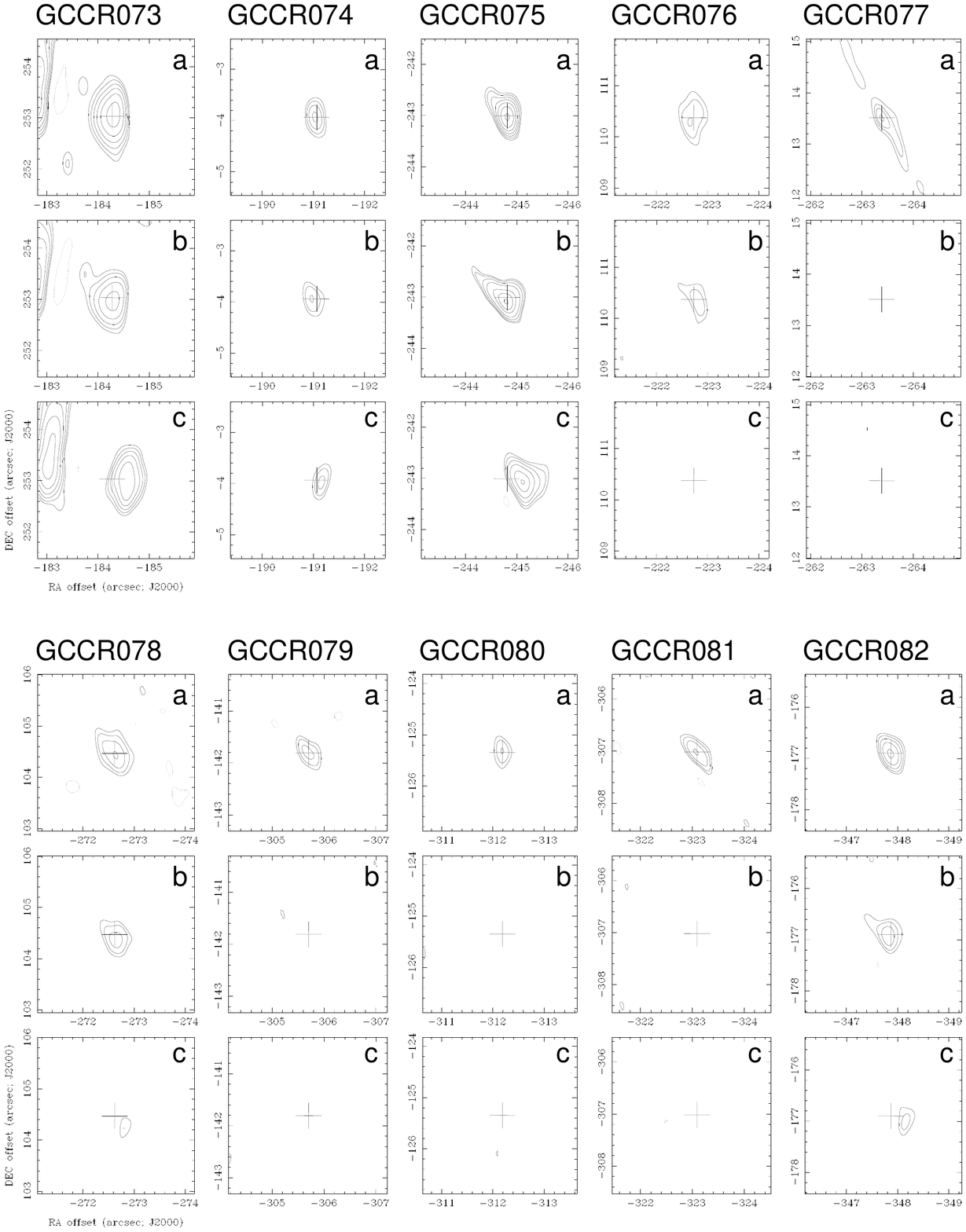}
\caption{Continued}
\label{fig:cont}
\end{figure*}

\begin{figure*}[tbh]
\centering
\includegraphics[angle=0,width=85mm]{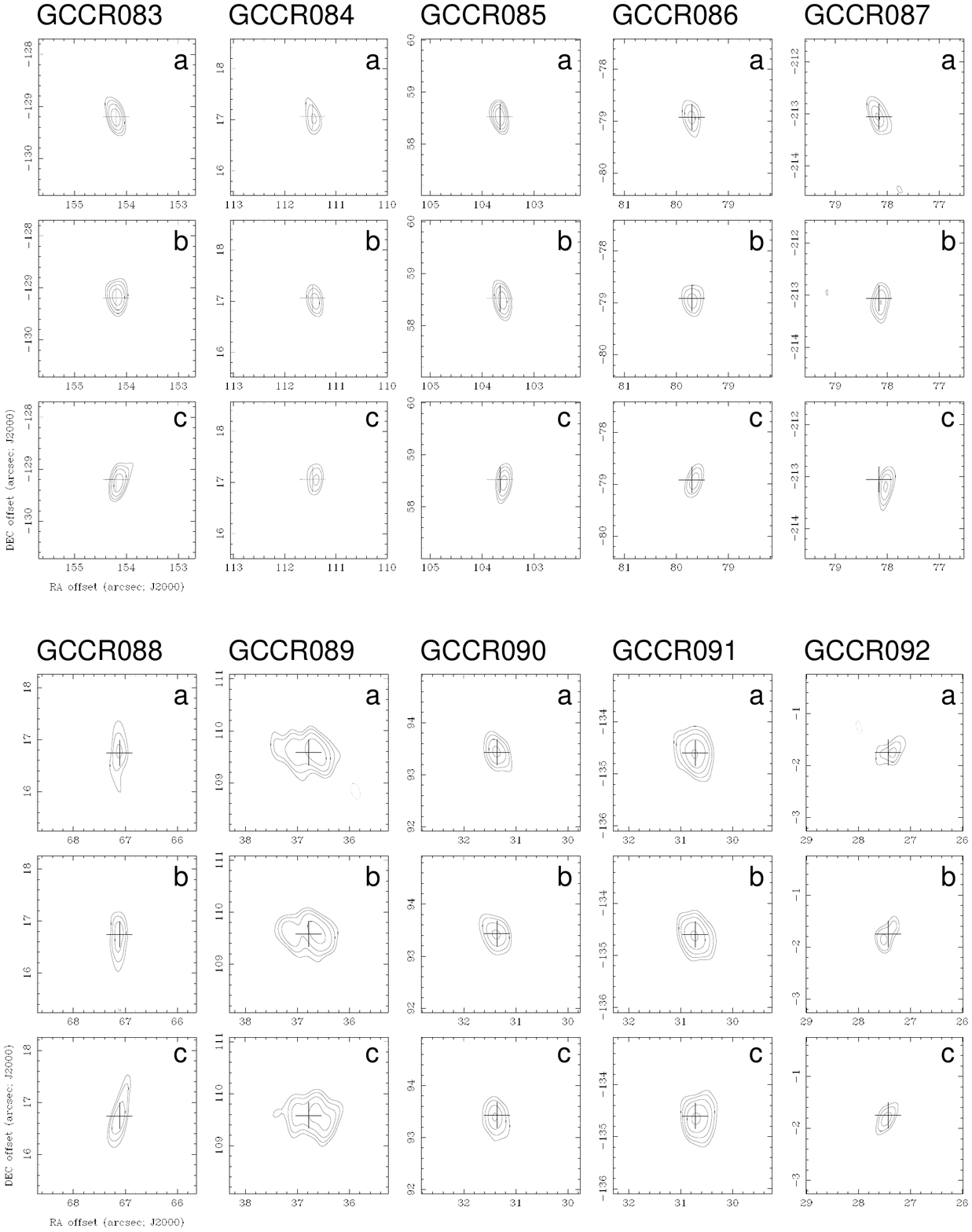}
\includegraphics[angle=0,width=85mm]{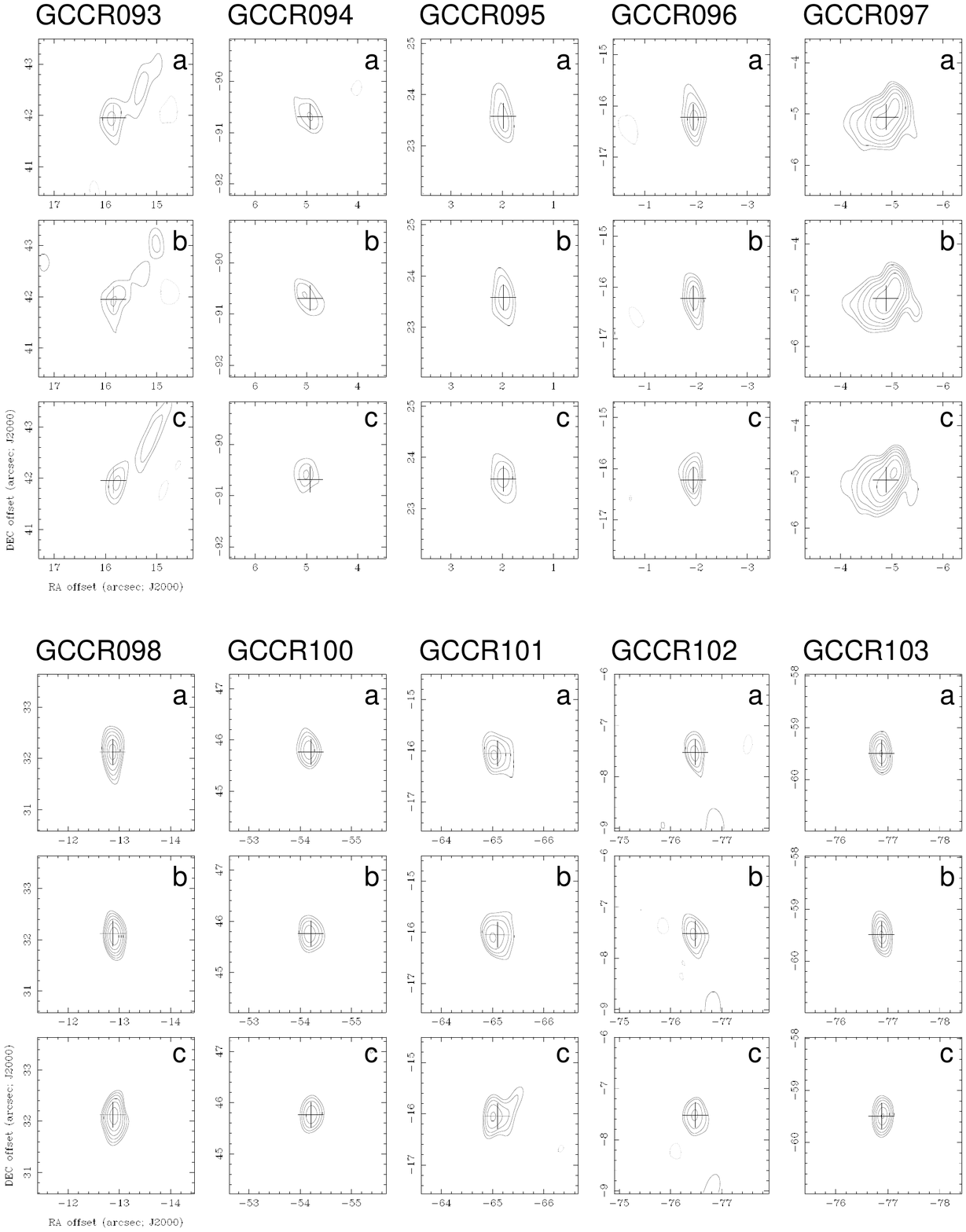} \\
\includegraphics[angle=0,width=143.6mm]{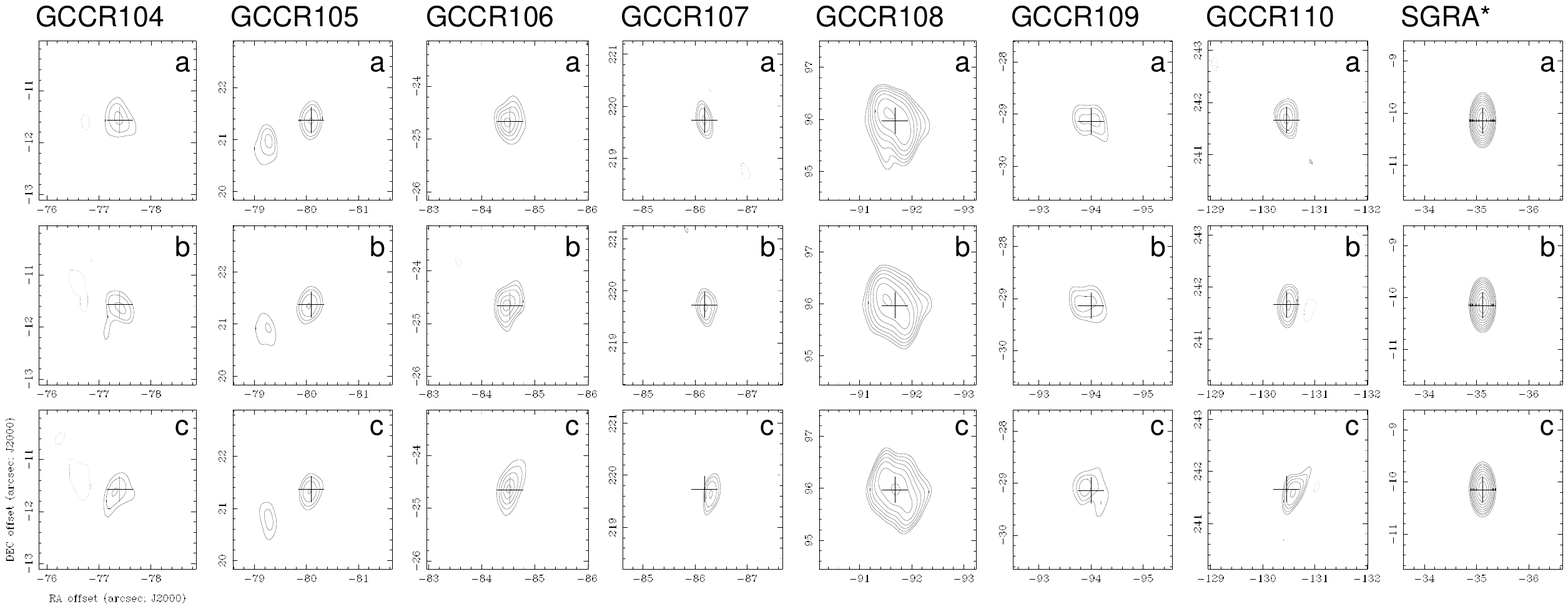}
\caption{Continued}
\label{fig:cont}
\end{figure*}

\begin{table*}[]
\footnotesize
\centering
\tablenum{2}
\setlength{\tabcolsep}{1.mm}
\caption{A catalog of compact radio sources at 5.5 GHz}
\begin{tabular} {cccrrccccccll}
\hline\hline \\
{ID}&
{RA(J2000) ~~ Dec(J2000)}&
{$r_\theta$}&
\multicolumn{1}{c}{$\Delta\alpha$}&
\multicolumn{1}{c}{$\Delta\delta$}&
{${\mathscr F}_{\rm PB}$}&
{$\sigma{\mathscr F}_{\rm PB}^{-1}$}&
{$S\pm\sigma$}&
{$S\pm\sigma$}&
{$S\pm\sigma$}&
{$\sigma_{\rm map}$} &
\multicolumn{2}{c}{Notes}\\
{} &
{} &
{(arcsec)}&
\multicolumn{2}{c}{(arcsec)}&
{} &
{} &
{(mJy)} &
{(mJy)} &
{(mJy)} &
{($\mu$Jy bm$^{-1}$)}&
{r-morph}&
{x-ID}\\
{(1)}&
{(2)}&
{(3)}&
\multicolumn{1}{c}{(4)}&
\multicolumn{1}{c}{(5)}&
{(6)}&
{(7)}&
{(8)}&
{(9)}&
{(10)}&
{(11)}&
{(12)}&
{(13)}\\
\hline \\ 
&&&&&&&\underline{2019-09-08}&\underline{2014-05-26}&\underline{2014-05-17}&\\
\multicolumn{9}{l}{\bf Variables and transients -}\\
GCCR001 &17:46:10.570  $-$28:55:49.08&453.8&400.6&279.0&42&6.6&17.2$\pm$0.8&7.29$\pm$0.3&$<$4.67&300&u-core&y\\
GCCR002 &17:46:06.132  $-$28:55:56.09&403.7&342.4&272.0&11&1.0&$<0.2$     &0.40$\pm$0.03&$<$0.31&35&u-core&n\\
GCCR003 &17:46:01.547  $-$29:01:04.21&251.4&282.1&$-$36.1&2.2&0.03&0.23$\pm$0.02&0.15$\pm$0.02&0.14$\pm$0.02&13&u-core&n\\
GCCR004 &17:45:58.370  $-$28:55:45.28&341.4&240.5&282.8&5.1&0.23&0.87$\pm$0.07&0.68$\pm$0.06&0.58$\pm$0.06&33&u-core&y\\
GCCR005 &17:45:57.717  $-$28:55:13.67&362.4&232.0&314.4&6.5&0.37&0.47$\pm$0.04&0.30$\pm$0.03&0.36$\pm$0.03&33&u-core&n\\
GCCR006 &17:45:58.348  $-$28:53:31.86&455.0&240.3&416.3&45&7.1&12.5$\pm$0.9&2.68$\pm$0.25&1.86$\pm$0.18&300&u-core&n\\
GCCR007 &17:45:57.931  $-$28:59:12.88&209.9&234.7&75.2&1.7&0.017&0.55$\pm$0.03&0.40$\pm$0.02&0.47$\pm$0.02&13&c-core$^1$&n\\
GCCR008 &17:45:57.165  $-$28:58:44.37&211.5&224.7&103.7&1.7&0.018&0.47$\pm$0.03&0.28$\pm$0.02&0.30$\pm$0.02&13&u-core&n\\
GCCR009 &17:45:57.117  $-$28:58:02.35&232.6&224.0&145.8&2.0&0.025&0.42$\pm$0.04&0.34$\pm$0.04&0.25$\pm$0.04&14&c-core$^2$&n\\
GCCR010 &17:45:57.107  $-$28:57:56.17&236.1&223.9&151.9&2.0&0.027&0.46$\pm$0.04&0.41$\pm$0.04&0.32$\pm$0.03&14&c-core$^3$&n\\
GCCR011 &17:45:55.082  $-$28:54:04.61&407.1&197.4&383.5&12&1.1&1.20$\pm$0.11&$<$0.05&$<$0.07&50&u-core&n\\
GCCR012 &17:45:53.901  $-$28:58:02.45&199.8&181.8&145.7&1.6&0.014&0.63$\pm$0.04&0.56$\pm$0.04&0.74$\pm$0.05&34&l-core$^4$&n\\
GCCR013 &17:45:50.764  $-$29:00:39.19&107.8&140.7&$-$11.1&1.1&0.004&0.09$\pm$0.01&0.05$\pm$0.01&0.06$\pm$0.01&13&u-core&n\\
GCCR014 &17:45:49.359  $-$29:04:42.47&278.5&122.2&$-$254.4&2.8&0.061&0.21$\pm$0.01&0.12$\pm$0.01&0.11$\pm$0.01&20&u-core&y\\
GCCR015 &17:45:49.330  $-$29:04:42.41&278.3&121.8&$-$254.3&2.7&0.060&0.19$\pm$0.01&0.15$\pm$0.01&0.16$\pm$0.01&20&u-core&y\\
GCCR016 &17:45:49.174  $-$28:59:33.27&95.8&119.8&54.8&1.1&0.004&0.18$\pm$0.01&0.15$\pm$0.01&0.14$\pm$0.01&7&t-core$^5$&n\\
GCCR017 &17:45:49.153  $-$28:59:30.03&97.1&119.5&58.1&1.1&0.004&0.17$\pm$0.01&$<$0.11&$<$0.10&7&d-core$^6$&n\\
GCCR018 &17:45:49.034  $-$29:00:19.57&82.9&118.0&8.5&1.1&0.004&0.10$\pm$0.02&$<$0.03&0.10$\pm$0.01&9&c-core$^7$&n\\
GCCR019 &17:45:48.908  $-$29:00:19.99&81.2&116.3&8.1&1.1&0.003&0.10$\pm$0.01& 0.08$\pm$0.01& 0.07$\pm$0.01&7&c-core$^8$&n\\
GCCR020 &17:45:48.837  $-$29:00:23.07&80.5&115.4&5.0&1.1&0.003&0.11$\pm$0.01&0.10$\pm$0.02&0.06$\pm$0.01&7&u-core&n\\
GCCR021 &17:45:48.711  $-$28:57:15.33&198.8&113.8&192.8&1.6&0.014&0.08$\pm$0.02&0.07$\pm$0.02&$<$0.10&7&u-core&n\\
GCCR022 &17:45:48.525  $-$29:00:37.25&78.6&111.3&$-$9.1&1.1&0.003&0.08$\pm$0.01&0.08$\pm$0.01&0.05$\pm$0.01&6&u-core&n\\
GCCR023 &17:45:48.508  $-$29:00:58.51&86.1&111.1&$-$30.4&1.1&0.004&0.08$\pm$0.01&0.10$\pm$0.01&0.04$\pm$0.01&7&u-core&n\\
GCCR024 &17:45:48.291  $-$28:59:25.02&90.2&108.2&63.1&1.1&0.004&0.09$\pm$0.01&$<$0.02&$<$0.03&7&u-core&n\\
GCCR025 &17:45:47.740  $-$29:00:52.12&74.2&101.0&$-$24.0&1.1&0.003&0.23$\pm$0.02&0.16$\pm$0.02&$<$0.05& 5& c-core$^9$&n\\
GCCR026 &17:45:47.696  $-$28:58:47.00&112.0&100.4&101.1&1.2&0.004&0.16$\pm$0.01&0.13$\pm$0.02&0.10$\pm$0.01&7&c-core$^{10}$&n\\
GCCR027 &17:45:46.300  $-$28:54:34.08&347.1&82.1&354.0&5.4&0.26&0.30$\pm$0.03&$<0.07$&$<0.06$&25&u-core&n\\
GCCR028 &17:45:45.638  $-$29:00:17.43&38.3&73.4&10.7&  1.0&0.003&0.14$\pm$0.01&0.10$\pm$0.01&0.12$\pm$0.01&10&c-core$^{11}$&n\\
GCCR029 &17:45:45.638  $-$29:00:22.31&38.5&73.4&5.8&1.0&0.003&0.31$\pm$0.01&0.31$\pm$0.01&0.40$\pm$0.02&12&l-core$^{12}$&n\\
GCCR030 &17:45:45.151  $-$29:00:37.45&37.4&67.0&$-$9.3&1.0&0.003&0.15$\pm$0.01&0.12$\pm$0.01&0.10$\pm$0.01&9&c-core$^{13}$&n\\
GCCR031 &17:45:44.932  $-$29:00:17.14&29.0& 64.2&11.0&1.0&0.003&0.10$\pm$0.01&0.07$\pm$0.01&0.06$\pm$0.01&6&c-core$^{14}$&y\\
GCCR032 &17:45:44.622  $-$29:01:23.93&70.5&60.1&$-$55.8&1.1&0.003&2.23$\pm$0.01&2.48$\pm$0.01&2.34$\pm$0.01&13&c-core$^{15}$&y\\
GCCR033 &17:45:44.500  $-$29:00:32.75&27.7&58.5&$-$4.6&1.0&0.003&0.19$\pm$0.02&0.13$\pm$0.01&0.09$\pm$0.01&7&c-core$^{16}$&n\\
GCCR034 &17:45:44.382  $-$28:59:07.15&74.1&56.9&81.0&1.1&0.003&0.07$\pm$0.01&0.14$\pm$0.01&0.10$\pm$0.01&7&c-core$^{17}$&n\\
GCCR035 &17:45:44.300  $-$29:00:17.56&20.8&55.9&10.6&1.0&0.003&0.12$\pm$0.01&$<$0.02&$<$0.02&10&u-core&y\\
GCCR036 &17:45:44.172  $-$29:01:27.87&72.5&54.2&$-$59.8&1.1&0.003&0.20$\pm$0.01&0.11$\pm$0.02&$<0.04$&8&c-core$^{18}$&y\\
GCCR037 &17:45:43.916  $-$29:00:21.27&16.0&50.8&6.8&1.0&0.003&0.25$\pm$0.02&0.17$\pm$0.02&0.17$\pm$0.02&10&c-core$^{19}$&y?\\
GCCR038 &17:45:43.036  $-$28:59:49.60&28.7&39.3&38.5&1.0&0.003&0.10$\pm$0.01&0.20$\pm$0.02&0.15$\pm$0.02&10&c-core$^{20}$&y?\\
GCCR039 &17:45:42.623  $-$29:00:24.80&6.9&34.3&3.3&1.0&0.003&0.17$\pm$0.02&0.22$\pm$0.02&0.30$\pm$0.02&10&c-core$^{21}$&n\\
GCCR040 &17:45:41.950  $-$29:01:00.78&44.0&25.0&$-$32.7&1.0&0.003&0.20$\pm$0.02&0.38$\pm$0.02&0.35$\pm$0.02&13&c-core$^{22}$&n\\
GCCR041 &17:45:41.664  $-$28:56:50.09&208.3& 21.3&218.0&1.7&0.017&0.22$\pm$0.02&$<$0.03&$<$0.03&10&c-core$^{23}$&y?\\
GCCR042 &17:45:41.522  $-$29:01:54.82&98.1&19.4&$-$86.7&1.1&0.004&0.13$\pm$0.01&0.09$\pm$0.01&0.08$\pm$0.01&10&u-core&n\\
GCCR043 &17:45:40.515  $-$29:05:03.65&287.1&6.2&$-$275.5&3.0&0.072&1.04$\pm$0.03&1.05$\pm$0.03&0.90$\pm$0.03&24&c-core$^{24}$&n\\
GCCR044 &17:45:39.474  $-$28:59:10.58&79.7&$-$7.4&77.5&1.1&0.003&0.49$\pm$0.02&0.50$\pm$0.02&0.38$\pm$0.02&18&c-core$^{25}$&n\\
GCCR045 &17:45:39.034  $-$29:01:43.24&98.0&$-$13.2&$-$75.1&1.1&0.004&1.00$\pm$0.02&1.19$\pm$0.02&1.21$\pm$0.02&11&t-core$^{26}$&n\\
GCCR046 &17:45:38.617  $-$28:59:03.21&92.1&$-$18.7&84.9&1.1&0.004&0.30$\pm$0.02&0.16$\pm$0.02&$<$0.05&13&c-core$^{27}$&n\\
GCCR047 &17:45:38.571  $-$29:01:36.75&95.7&$-$19.3&$-$68.6&1.1&0.004&0.32$\pm$0.02&0.34$\pm$0.02&0.29$\pm$0.01&13&c-core$^{28}$&y\\
GCCR048 &17:45:38.358  $-$29:01:48.19&106.8&$-$22.1&$-$80.1&1.1&0.004&0.26$\pm$0.01&0.19$\pm$0.01&0.21$\pm$0.01&9&c-core$^{29}$&n\\
GCCR049 &17:45:37.958  $-$29:01:47.05&108.8&$-$27.3&$-$78.9&1.1&0.004&0.28$\pm$0.02&0.37$\pm$0.03&0.28$\pm$0.02&13&c-core$^{30}$&y\\
GCCR050 &17:45:37.850  $-$29:00:26.15&64.4&$-$28.7&2.0& 1.0&0.003&0.66$\pm$0.01&0.97$\pm$0.02&0.94$\pm$0.02&22&c-core$^{31}$&y\\
GCCR051 &17:45:37.753  $-$28:57:15.07&194.2&$-$30.0&193.0&1.6&0.013&1.23$\pm$0.05&0.92$\pm$0.04&1.15$\pm$0.04&23&l-core$^{32}$&n\\
GCCR052 &17:45:37.463  $-$29:00:14.55&69.1&$-$33.8&13.6&1.1&0.003&0.89$\pm$0.02&0.71$\pm$0.02&0.97$\pm$0.02&29&c-core$^{33}$&n\\
GCCR053 &17:45:37.390  $-$28:59:23.23&88.8&$-$34.8&64.9&1.1&0.004&0.29$\pm$0.02&0.22$\pm$0.02&0.18$\pm$0.01&12&t-core$^{34}$&n\\
GCCR054 &17:45:37.375  $-$29:00:32.61&71.5&$-$35.0&$-$4.5&1.1&0.003&1.54$\pm$0.02&1.95$\pm$0.02&1.82$\pm$0.02&35&l-core$^{35}$&n\\
GCCR055 &17:45:37.310  $-$29:00:31.98&72.3&$-$35.8&$-$3.9&1.1&0.003&0.84$\pm$0.02&0.78$\pm$0.02&0.56$\pm$0.01&35&d-core$^{36}$&n\\
GCCR056 &17:45:36.920  $-$29:00:39.17&78.9&$-$40.9&$-$11.1&1.1&0.003&8.16$\pm$0.03&8.08$\pm$0.03&8.40$\pm$0.03&110&t-core$^{37}$&y\\
GCCR057 &17:45:36.425  $-$29:00:43.37&86.3&$-$47.4&$-$15.3&1.1&0.004&0.66$\pm$0.02&0.62$\pm$0.02&0.54$\pm$0.02&13&t-core$^{38}$&y\\
GCCR058 &17:45:35.804 $-$29:00:04.13&91.6&$-$55.6&24.0 &1.1&0.003&1.28$\pm$0.02&1.49$\pm$0.02&1.49$\pm$0.02&14&l-core$^{39}$&y\\
GCCR059 &17:45:35.499 $-$28:59:53.83&97.7&$-$59.6&34.3 &1.1&0.004&0.17$\pm$0.02&0.23$\pm$0.02&0.25$\pm$0.02&11&c-core$^{40}$&y?\\
GCCR060 &17:45:34.271 $-$29:03:36.62&227.4&$-$75.7&$-$188.5&1.9&0.023&0.13$\pm$0.01&$<$0.08&$<$0.09&10&c-core$^{41}$&n\\
GCCR061 &17:45:34.068 $-$29:00:29.91&114.0&$-$78.4&$-$1.8&1.2&0.004&0.24$\pm$0.02&0.15$\pm$0.02&0.19$\pm$0.02&9&l-core$^{42}$&y\\
GCCR062 &17:45:33.867 $-$28:57:43.58&193.1&$-$81.0&164.5&1.6&0.013&0.24$\pm$0.02&0.12$\pm$0.02&$<$0.05&9&c-core$^{43}$&n\\
GCCR063 &17:45:33.749 $-$28:55:30.85&310.3&$-$82.6&297.3&3.7&0.11&0.82$\pm$0.02&0.60$\pm$0.02&0.58$\pm$0.02&20&l-core$^{44}$&n\\
\\
\hline
\end{tabular}
\end{table*}

\begin{table*}[]
\footnotesize
\centering
\setlength{\tabcolsep}{1.mm}
\tablenum{2}\caption{$-$ Continued}
\begin{tabular} {cccrrcccccccll}
\hline\hline \\
{ID}&
{RA(J2000) ~~ Dec(J2000)}&
{$r_\theta$}&
\multicolumn{1}{c}{$\Delta\alpha$}&
\multicolumn{1}{c}{$\Delta\delta$}&
{${\mathscr F}_{\rm PB}$}&
{$\sigma{\mathscr F}_{\rm PB}^{-1}$}&
{$S\pm\sigma$}&
{$S\pm\sigma$}&
{$S\pm\sigma$}&
{$\sigma_{\rm map}$} &
\multicolumn{2}{c}{Notes}\\
{} &
{} &
{(arcsec)}&
\multicolumn{2}{c}{(arcsec)}&
{} &
{} &
{(mJy)} &
{(mJy)} &
{(mJy)} &
{($\mu$Jy bm$^{-1}$)}&
{r-morph}&
{x-ID}\\
{(1)}&
{(2)}&
{(3)}&
\multicolumn{1}{c}{(4)}&
\multicolumn{1}{c}{(5)}&
{(6)}&
{(7)}&
{(8)}&
{(9)}&
{(10)}&
{(11)}&
{(12)}&
{(13)} \\
\hline \\ 
&&&&&&&\underline{2019-09-08}&\underline{2014-05-26}&\underline{2014-05-17}&\\
GCCR064 &17:45:33.610 $-$29:01:40.76&145.4&$-$84.4&$-$72.6&1.3&0.006&0.66$\pm$0.02&0.55$\pm$0.02&0.60$\pm$0.02&16&t-core$^{45}$&y\\
GCCR065 &17:45:32.927 $-$28:56:11.34&278.1&$-$~93.4&256.8&2.7&0.060&0.26$\pm$0.02&0.14$\pm$0.02&$<$0.10&19&c-core$^{46}$&n\\
GCCR066 &17:45:32.767 $-$28:56:10.82&279.5&$-$~95.5&257.3&2.8&0.062&0.35$\pm$0.02&0.25$\pm$0.02&0.20$\pm$0.02&22&u-core&n\\
GCCR067 &17:45:32.613 $-$29:00:42.60&134.9&$-$~97.4&$-$14.5&1.2&0.006&0.15$\pm$0.01&0.08$\pm$0.01&0.08$\pm$0.01&10&c-core$^{47}$&y\\
GCCR068 &17:45:32.552 $-$29:00:21.95&133.4&$-$~98.3&6.2&1.2&0.005&0.18$\pm$0.01&0.19$\pm$0.01&0.14$\pm$0.01&9&u-core&n\\
GCCR069 &17:45:32.025 $-$28:56:18.70&277.4&$-$105.2&249.4&2.7&0.059&0.82$\pm$0.03&0.70$\pm$0.03&0.60$\pm$0.03&24&l-core$^{48}$&n\\
GCCR070 &17:45:30.031 $-$28:59:42.17&170.3&$-$131.3&45.9&1.4&0.009&1.28$\pm$0.03&1.11$\pm$0.03&1.12$\pm$0.03&12&t-core$^{49}$&y\\
GCCR071 &17:45:29.947 $-$28:54:20.64&394.7&$-$132.5&367.5&10&0.83&1.39$\pm$0.12&$<0.16$&$<0.20$&48&t-core$^{50}$&n\\
GCCR072 &17:45:28.892 $-$28:57:26.02&249.9&$-$146.3&182.1&2.2&0.035&0.23$\pm$0.02&0.15$\pm$0.01&0.17$\pm$0.01&13&u-core&y\\
GCCR073 &17:45:28.671 $-$28:56:04.94&313.0&$-$149.2&263.2&3.8&0.12&8.78$\pm$0.03&5.85$\pm$0.03&7.12$\pm$0.03&100&c-core$^{51}$&y\\
GCCR074 &17:45:28.154 $-$29:00:21.91&190.9&$-$155.9&6.2&1.6&0.012&0.18$\pm$0.01&0.13$\pm$0.01&0.12$\pm$0.01&10&u-core&y?\\
GCCR075 &17:45:24.057 $-$29:04:20.98&344.8&$-$209.6&$-$232.9&5.3&0.24&1.80$\pm$0.05&2.20$\pm$0.09&1.76$\pm$0.09&40&u-core&n\\
GCCR076 &17:45:25.740 $-$28:58:27.60&248.5&$-$187.6&120.5&2.2&0.034&0.36$\pm$0.03&0.25$\pm$0.02&0.22$\pm$0.02&13&t-core$^{52}$&n\\
GCCR077 &17:45:22.641 $-$29:00:04.46&263.7&$-$228.3&~23.7&2.5&0.045&0.20$\pm$0.02&$<$0.06&$<$0.06&15&l-core$^{53}$&y?\\
GCCR078 &17:45:21.937 $-$28:58:33.50&292.0&$-$237.5&114.6&3.1&0.079&0.44$\pm$0.02&0.27$\pm$0.02&0.16$\pm$0.01&14&c-core$^{54}$&n\\
GCCR079 &17:45:19.415 $-$29:02:39.78&337.0&$-$270.5&$-$131.7&4.8&0.20&0.25$\pm$0.02&$<$0.08&$<$0.1&20&u-core&n\\
GCCR080 &17:45:18.921 $-$29:02:23.32&336.4&$-$277.0&$-$115.2&4.8&0.20&0.21$\pm$0.02&$<$0.07&$<$0.08&20&u-core&n\\
GCCR081 &17:45:18.090 $-$29:05:24.99&445.6&$-$287.9&$-$296.9&30&4.3&1.93$\pm$0.19&0.80$\pm$0.19&$<$0.06&120&u-core&n\\
GCCR082 &17:45:16.201 $-$29:03:14.87&390.1&$-$312.7&$-$166.8&9.5&0.74&1.22$\pm$0.05&0.98$\pm$0.05&0.61$\pm$0.06&50&u-core&y\\
\multicolumn{8}{l}{~}\\
\hline \\
&&&&&&&\underline{2019-09-08}&\underline{2014-05-26}&\underline{2014-05-17}\\
\multicolumn{8}{l}{\bf Non-variables -} \\
GCCR083 &17:45:54.472  $-$29:02:27.17&201.2&189.3&$-$119.1&1.6&0.015&0.16$\pm0$.01&0.17$\pm$0.01&0.16$\pm$0.01&10&u-core&n\\
GCCR084 &17:45:51.214  $-$29:00:00.91&112.9&146.6&27.2&1.2&0.004&0.10$\pm$0.01&0.11$\pm$0.01&0.10$\pm0.01$&10&u-core&n\\
GCCR085 &17:45:50.620  $-$28:59:19.45&119.0&138.8&68.7&1.2&0.005&0.10$\pm0.01$&0.11$\pm$0.01&0.09$\pm$0.01&7&u-core&y\\
GCCR086 &17:45:48.794  $-$29:01:36.89&112.2&114.8&$-$68.8&1.2&0.004&0.12$\pm$0.01&0.14$\pm$0.01&0.12$\pm$0.01&10&u-core&n\\
GCCR087 &17:45:48.676  $-$29:03:51.03&226.9&113.2&$-$202.9&1.9&0.023&0.15$\pm$0.01&0.14$\pm$0.01&0.14$\pm$0.01&11&u-core&y\\
GCCR088 &17:45:47.834  $-$29:00:01.23&69.2&102.2&26.9&1.1&0.003&0.22$\pm$0.02&0.24$\pm$0.02&0.25$\pm$0.02&10&l-core$^{55}$&n\\
GCCR089 &17:45:45.522  $-$28:58:28.39&115.6&71.9&119.7&1.2&0.005&0.54$\pm$0.02&0.55$\pm$0.02&0.53$\pm$0.02&9&l-core$^{56}$&y\\
GCCR090 &17:45:45.109  $-$28:58:44.54&98.5&66.5&103.6&1.1&0.004&0.20$\pm$0.01&0.21$\pm$0.01&0.21$\pm$0.01&9&c-core$^{57}$&n\\
GCCR091 &17:45:45.060  $-$29:02:32.57&138.1&65.8&$-$124.5&1.3&0.006&0.47$\pm$0.02&0.44$\pm$0.02&0.45$\pm$0.02&10&c-core$^{58}$&n\\
GCCR092 &17:45:44.805  $-$29:00:19.72&27.6&62.5&8.4&1.0&0.003&0.09$\pm$0.01&0.10$\pm$0.01&0.10$\pm$0.01&8&c-core$^{59}$&y?\\
GCCR093 &17:45:43.926  $-$28:59:36.02&44.8&51.0&52.1&1.0&0.003&0.24$\pm$0.02&0.20$\pm$0.02&0.19$\pm$0.02&10&t-core$^{60}$&n\\
GCCR094 &17:45:43.094  $-$29:01:48.66&90.8&40.0&$-$80.5&1.1&0.004&0.21$\pm$0.02&0.20$\pm$0.02&0.22$\pm$0.02&10&c-core$^{61}$&n\\
GCCR095 &17:45:42.869  $-$28:59:54.39&23.7&37.1&33.7&1.0&0.003&0.55$\pm$0.02&0.55$\pm$0.02&0.55$\pm$0.02&20&l-core$^{62}$&y?\\
GCCR096 &17:45:42.570  $-$29:00:34.19&16.3&33.2&$-$6.1&1.0&0.003&0.29$\pm$0.02&0.27$\pm$0.02&0.27$\pm$0.02&10&l-core$^{63}$&y?\\
GCCR097 &17:45:42.346  $-$29:00:23.03&7.0&30.2&5.1&1.0&0.003&2.24$\pm$0.07&2.41$\pm$0.07&2.44$\pm$0.07&20&t-core$^{64}$&y?\\
GCCR098 &17:45:41.737  $-$28:59:45.85&34.6&22.2&42.3&1.0&0.003&0.28$\pm$0.02&0.33$\pm$0.02&0.28$\pm$0.02&10&c-core$^{65}$&y\\
GCCR099 &17:45:38.610  $-$29:01:35.31&94.3&$-$18.8&$-$67.2&1.1&0.004&0.40$\pm$0.04&0.40$\pm$0.04&0.43$\pm$0.04&13&l-core$^{66}$&y?\\
GCCR100 &17:45:38.586  $-$28:59:32.21&70.9&$-$19.1&55.9&1.1&0.003&0.23$\pm$0.01&0.24$\pm$0.01&0.23$\pm$0.01&12&c-core$^{67}$&y?\\
GCCR101 &17:45:37.756  $-$29:00:34.02&67.1&$-$30.0&$-$5.9&1.1&0.003&0.39$\pm$0.02&0.44$\pm$0.02&0.39$\pm$0.02&12&t-core$^{68}$&y?\\
GCCR102 &17:45:36.888  $-$29:00:25.49&76.8&$-$41.4&2.6&1.1&0.003&0.24$\pm$0.02&0.23$\pm$0.02&0.26$\pm$0.02&11&c-core$^{69}$&n\\
GCCR103 &17:45:36.858  $-$29:01:17.46&97.2&$-$41.8&$-$49.3&1.1&0.004&0.24$\pm$0.02&0.22$\pm$0.01&0.22$\pm$0.01&10&u-core&y?\\
GCCR104 &17:45:36.818  $-$29:00:29.54&78.2&$-$42.3&$-$1.4&1.1&0.003&0.40$\pm$0.02&0.41$\pm$0.02&0.40$\pm$0.02&16&c-core$^{70}$&n\\
GCCR105 &17:45:36.613  $-$28:59:56.60&82.9&$-$45.0&31.5&1.1&0.004&0.30$\pm$0.02&0.34$\pm$0.02&0.32$\pm$0.02&20&c-core$^{71}$&n\\
GCCR106 &17:45:36.274  $-$29:00:42.63&88.0&$-$49.4&$-$14.5&1.1&0.004&0.22$\pm$0.02&0.23$\pm$0.02&0.22$\pm$0.02&9&c-core$^{72}$&y?\\
GCCR107 &17:45:36.149  $-$28:56:38.24&236.0&$-$51.1&229.9&2.0&0.027&0.09$\pm$0.01&0.12$\pm$0.01&0.10$\pm$0.01&9&u-core&y\\
GCCR108 &17:45:35.730  $-$28:58:42.00&132.8&$-$56.6&106.1&1.2&0.005&1.51$\pm$0.04&1.58$\pm$0.04&1.58$\pm$0.04&10&l-core$^{73}$&n\\
GCCR109 &17:45:35.553  $-$29:00:47.11&98.4&$-$58.9&$-$19.0&1.1&0.004&0.23$\pm$0.03&0.27$\pm$0.03&0.23$\pm$0.03&10&l-core$^{74}$&n\\
GCCR110 &17:45:32.758  $-$28:56:16.37&274.6&$-$95.4&251.8&2.7&0.056&0.23$\pm$0.02&0.21$\pm$0.02&0.23$\pm$0.03&11&u-core&y\\
\\
\hline
\end{tabular}
\end{table*}
\begin{table*}[ht!]
\footnotesize
\centering
\tablenum{2}\caption{$-$ Continued}
\begin{tabular}{llll}
\hline \\
\\
\multicolumn{4}{l}{\footnotesize Note to Table 2 $-$ Listed below are the source sizes 
{$\theta_{\rm maj}\pm\sigma,\theta_{\rm min}\pm\sigma$, PA$_\Theta\pm\sigma$} in the units of
(arcsec$,$arcsec, deg):} \\
\\
$^1$0.20$\pm0.02,0.12\pm0.01$, 17$\pm$5;&
$^2$0.63$\pm0.10,0.29\pm0.10$, 34$\pm$11;&
$^3$0.67$\pm0.12,0.31\pm0.11$, 18$\pm$10;&
$^4$0.96$\pm0.05,0.10\pm0.09$, 148$\pm$5;\\
$^5$0.75$\pm0.20,0.31\pm0.08$, 10$\pm$13;&
$^6$0.71$\pm0.10,0.24\pm0.16$, 108$\pm$12;&
$^7$0.50$\pm0.20,0.30\pm$0.17, 50$\pm$30;&
$^8$0.57$\pm0.12,0.1\pm0.1$, 71$\pm$15;\\
$^9$0.65$\pm0.10,0.28\pm$0.10,  130$\pm$10;&
$^{10}$0.46$\pm0.1,0.20\pm0.08$, 85$\pm$21;&
$^{11}$0.32$\pm0.05,0.04\pm$0.03, 15$\pm$4;&
$^{12}$0.80$\pm0.10,0.21\pm$0.04, 164$\pm$3;\\
$^{13}$0.50$\pm0.10,0.17\pm$0.08, 20$\pm$13;&
$^{14}$0.35$\pm0.13,0.26\pm$0.12, 68$\pm$21;&
$^{15}$0.37$\pm0.03,0.28\pm$0.03, 160$\pm$3;&
$^{16}$0.51$\pm0.07,0.24\pm$0.03,  $-4\pm$4;\\
$^{17}$0.28$\pm0.04,0.23\pm$0.02, 22$\pm$5;&
$^{18}$0.61$\pm0.16,0.28\pm$0.06,  13$\pm$8;&
$^{19}$0.77$\pm0.08,0.26\pm$0.07,  20$\pm$6;&
$^{20}$0.50$\pm0.10,0.33\pm$0.06,  174$\pm$25;\\
$^{21}$0.40$\pm0.08, 0.22\pm$0.07,  35$\pm$7;&
$^{22}$0.55$\pm0.10,0.25\pm$0.05,  4$\pm$8;&
$^{23}$0.41$\pm0.12,0.11\pm$0.06, 22$\pm$10;&
$^{24}$0.47$\pm0.08,0.14\pm0.05$, 41$\pm$8;\\
$^{25}$0.68$\pm0.09,0.43\pm$0.07, 67$\pm$15;&
$^{26}$0.40$\pm0.03,0.37\pm$0.03, 146$\pm$19;&
$^{27}$0.53$\pm0.21,0.14\pm$0.20, 57$\pm$23;&
$^{28}$0.65$\pm0.07,0.12\pm$0.08, 41$\pm$5;\\
$^{29}$0.35$\pm0.07,0.16\pm$0.04, 20$\pm$10;&
$^{30}$0.98$\pm0.05,0.05\pm$0.05, 144$\pm$6;&
$^{31}$0.39$\pm0.05,0.21\pm$0.02, 5$\pm$15;&
$^{32}$0.74$\pm0.10,0.45\pm$0.05, 64$\pm$6;\\
$^{33}$0.55$\pm0.05,0.41\pm$0.03, 49$\pm$8;&
$^{34}$0.46$\pm0.07,0.33\pm$0.04, 170$\pm$15;&
$^{35}$0.55$\pm0.05,0.38\pm$0.03, 80$\pm$5;&
$^{36}$0.59$\pm0.04,0.34\pm$0.03, 124$\pm$5;\\
\multicolumn{4}{l}{$^{37}$0.62$\pm0.04,0.42\pm$0.03, 78$\pm$15, a core of  M source \citep{yusef1987,zmg2013};}\\
$^{38}$0.63$\pm0.03,0.31\pm$0.04, 43$\pm$5;&
$^{39}$0.62$\pm0.04,0.31\pm$0.03, 90$\pm$4;&
$^{40}$0.61$\pm0.10,0.43\pm$0.10, 138$\pm$16;&
$^{41}$0.50$\pm0.20,0.20\pm$0.15, 134$\pm$30;\\
$^{42}$0.90$\pm0.10,0.25\pm$0.05, 176$\pm$4;&
$^{43}$0.70$\pm0.10,0.37\pm$0.10, 32$\pm$12;&
$^{44}$0.54$\pm0.05,0.37\pm$0.07, 97$\pm$10;&
$^{45}$0.76$\pm0.11,0.46\pm$0.07, 35$\pm$10;\\
$^{46}$0.28$\pm0.05,0.14\pm$0.13, 63$\pm$20;&
$^{47}$0.39$\pm0.10,0.23\pm$0.10, 40$\pm$25;&
$^{48}$0.70$\pm0.05,0.36\pm$0.05, 88$\pm$5;&
$^{49}$0.83$\pm0.05,0.51\pm$0.03, 154$\pm$5;\\
$^{50}$0.65$\pm0.15,0.10\pm$0.10, 99$\pm$20;&
\multicolumn{3}{l}{$^{51}$0.72$\pm0.05,0.35\pm$0.05, 5$\pm$3, source H2 \citep{yusef1987,zha93};}\\
$^{52}$0.69$\pm0.25,0.30\pm$0.07, 20$\pm$10;&
$^{53}$0.77$\pm0.21,0.38\pm$0.11, 29$\pm$21;&
$^{54}$0.43$\pm0.07,0.28\pm$0.07, 67$\pm$19;&
$^{55}$0.91$\pm0.10,0.20\pm0.05$, 172$\pm$5;\\
\multicolumn{4}{l}{$^{56}$0.85$\pm0.06,0.31\pm0.05$, 84$\pm$5, the core of Cannonball\citep{zmg2013};}\\
$^{57}$0.43$\pm0.06,0.12\pm$0.03, 42$\pm$7;&
$^{58}$0.49$\pm0.10,0.45\pm$0.13, 60$\pm$20;&
$^{59}$0.36$\pm0.13,0.18\pm$0.08, 132$\pm$20;&
$^{60}$0.63$\pm0.17,0.32\pm$0.07,  150$\pm$15;\\
$^{61}$0.63$\pm0.12,0.23\pm$0.06,  45$\pm$10;&
$^{62}$0.75$\pm0.10,0.36\pm$0.07,  15$\pm$5;&
$^{63}$0.60$\pm0.05,0.20\pm$0.05,  12$\pm$4;&
$^{64}$0.81$\pm0.03,0.44\pm$0.03,  133$\pm$3;\\
$^{65}$0.42$\pm0.03,0.11\pm$0.03, 4$\pm$3;&
$^{66}$one of the pair GCCR047/099;&
$^{67}$0.27$\pm0.05,0.15\pm$0.03, 47$\pm$6;&
$^{68}$0.44$\pm0.05,0.29\pm$0.07, 122$\pm$12;\\
$^{69}$0.43$\pm0.09,0.18\pm$0.04, 17$\pm$6;&
$^{70}$0.60$\pm0.10,0.47\pm$0.06, 33$\pm$14;&
$^{71}$0.24$\pm0.04,0.14\pm$0.07, 136$\pm$22;&
$^{72}$0.30$\pm0.06,0.29\pm$0.08, 98$\pm$30;\\
$^{73}$0.80$\pm$0.06,0.45$\pm$0.04,50$\pm$10;&
$^{74}$0.54$\pm0.10\times0.12\pm$0.10, 67$\pm$20;
\end{tabular}
\end{table*}
\twocolumn